\documentstyle[11pt]{article}

\newlength{\dinwidth}
\newlength{\dinmargin}
\def\slepton{\widetilde \ell}

\def\sll{\slepton_{L}}
\def\squark{\widetilde q}

\def\wino{{\widetilde{W}}}

\def\gluino{\widetilde g}

\def\sneu{\widetilde \nu}
\def\sele{\widetilde e}
\def\sqk{\widetilde q}
\def\ser{\sele_{R}}

\def\sql{\squark_{L}}

\def\su{\widetilde{u}}
\def\sd{\widetilde{d}}

\def\sur{\su_{R}}
\def\sdr{\sd_{R}}

\def\st{\widetilde{t}}

\def\ddf{{\rm d}}
\def\bino{\widetilde{B}}

\def\sz1{{\widetilde{Z}}_{1}}

\def\szk{{\widetilde{Z}}_{k}}
\def\swl{{\widetilde{W}}_{1}}

\def\swi{{\widetilde{W}}_{i}}

\def\msz1{m_{\sz1}}

\def\tanbe{\tan\beta}
\def\nle{{\stackrel{<}{\sim}}}
\def\nge{{\stackrel{>}{\sim}}}

\def\half{{\frac{1}{2}}}
\def\mt{m_{t}}

\def\stl{\st_{1}}
\def\stls{\st^{*}_{1}}
\def\sth{\st_{2}}

\def\mstl{m_{\stl}}

\def\mz{m_{Z}}

\def\tht{\theta_{t}}

\def\tew{\theta_{W}}

\def\sw{\sin^{2}\theta_{W}}

\def\mtil{\widetilde{m}}

\def\rb{\slash\hspace{-9pt}R}

\def\msh1{\mtil^{2}_{H_{1}}}
\def\msh2{\mtil^{2}_{H_{2}}}

\def\frb{F_{\rm RB}}
\def\i{{\rm i}}
\def\xq{(x, Q^{2})}
\def\tui{T^{u}_{\i}}
\def\tuib{\bar{T}^{u}_{\i}}
\def\tdi{T^{d}_{\i}}
\def\tdib{\bar{T}^{d}_{\i}}
\def\tuo{T^{u}_{1}}
\def\tuob{\bar{T}^{u}_{1}}
\def\tdo{T^{d}_{1}}
\def\tdob{\bar{T}^{d}_{1}}
\def\tut{T^{u}_{2}}
\def\tutb{\bar{T}^{u}_{2}}
\def\tdt{T^{d}_{2}}
\def\tdtb{\bar{T}^{d}_{2}}
\def\tdh{T^{d}_{3}}
\def\tdhb{\bar{T}^{d}_{3}}
\def\tdf{T^{d}_{4}}
\def\tdfb{\bar{T}^{d}_{4}}
\def\ee{e^{e}}
\def\eu{e^{u}}
\def\ed{e^{d}}
\def\mz{m_{Z}}
\def\ael{A^{e}_{L}}
\def\aer{A^{e}_{R}}

\def\adr{A^{d}_{R}}
\def\aul{A^{u}_{L}}
\def\aur{A^{u}_{R}}
\def\qz{{{Q^2}\over{Q^2 +\mz^2}}}
\def\utg{(Q^2-sx-\mstl^{2})}
\def\stg{(sx-\mstl^{2})^{2}+\mstl^{2}\Gamma_{\stl}^{2}}
\begin{document}
{}~~~\\
\vspace{10mm}
\begin{flushright}
ITP-SU-95/03 \\
hep-ph/9509294\\
\vspace{0.3cm}
September, 1995
\end{flushright}
\begin{center}
  \begin{Large}
   \begin{bf}
Supersymmetric Particle Production at HERA
 \\
   \end{bf}
  \end{Large}
  \vspace{5mm}
  \begin{large}
    Tetsuro Kobayashi\\
  \end{large}
Faculty of Engineering,
Fukui Institute of Technology, Fukui 910, Japan \\
koba@ge.seikei.ac.jp\\
  \vspace{3mm}
 \begin{large}
    Shoichi Kitamura \\
 \end{large}
Tokyo Metropolitan College of Allied Medical Sciences, Tokyo 116, Japan\\
    kitamura@post.metro-ms.ac.jp\\
 \vspace{3mm}
 \begin{large}
Tadashi Kon \\
  \end{large}
Faculty of Engineering, Seikei University, Tokyo 180, Japan \\
kon@ge.seikei.ac.jp\\
  \vspace{5mm}
\end{center}
\vskip50pt
\begin{quotation}
\noindent
\begin{center}
{\bf Abstract}
\end{center}
 In the framework of the minimal supersymmetric standard model
and the $R$-parity breaking model,
we investigate various
production processes
of the supersymmetric partner at HERA energies.
Our emphasis is paid upon the scalar top quark, the partner of
top quark, characterized by its lighter mass than the top quark and other
scalar quarks in a model.
We propose experimentally feasible approaches to search for
clean signals of the stop from either its production or decay processes.
\end{quotation}
\vfill\eject
\section{\it Introduction}

   It is very amazing that the Standard Model (SM) has been so successful
\cite{dp}.
Discoveries of $W^{\pm}$ and $Z^0$ and high level precision electroweak
tests by LEP have been vindicating the SM over and over again.
The discovery of the top quark at TEVATRON \cite{top} seems to be its
culmination.
The theoretical prediction agrees very well with experiments.
However, the problems of $R_b$ and $R_c$ may shed a light on the
fate of the SM \cite{Hagiwara}.
As far as the first approximation is concerned there is no need to go
beyond the SM.
However, the SM cannot be a fundamental theory in any sense.
Actually, there are a large number of free
parameters, the arbitrariness of particle masses and mixing
angles, and the lack of any explanation for the replication of
generations and so on.
Moreover, it is known that the gauge hierarchy problem exists in the SM.
Theorists have so far attempted to find a way to go beyond the SM via
various standpoints :
unification \cite{unif}, technicolor \cite{tech}, supersymmetry (SUSY)
\cite{Nilles}
or superstring \cite{sst}.

 Among them the SUSY seems to be the most reasonable candidate to our
 expectation.
 A number of theoretical and phenomenological reasons make low energy SUSY
 attractive with respect to its alternative.
 The supersymmetry is a symmetry between fermions and bosons.
 In the SUSY models, the quadratic divergence can be cancelled out owing to
both contributions of boson and fermion loops.
This is a key point for the fact that the hierarchy problem can be
solved in the SUSY models.
However, we should note that fermions and bosons,
in other words, the ordinary particles and their SUSY partners (sparticles),
must be degenerate in mass in the exact SUSY limit.
Obviously there appears to be no
evidence in nature for such a situation.
Therefore, in order to apply SUSY to
particle physics, we must consider models in which SUSY is broken.
In this case masses of all sparticles must be less than about
1 TeV in order to solve the hierarchy problem.
At first sight, it seems to be unnatural that we need a large number of
new particles, sparticles, undiscovered yet.
However, such new particles play an essential role in
some phenomenologically favourable properties in the model.
The most impressive evidence in favour of SUSY may be the unification of
gauge couplings in SUSY Grand Unified Theories (GUTs) \cite{running}.
The dark matter issue \cite{dm} in astrophysics also encourages SUSY
proponents.
Most of us consider the "SUSY world" as a plausible scenario for future
particle physics.

  HERA at DESY is the world first electron-proton collider
whose $ep$ center of mass energy is 314 GeV produced by 30 GeV
electron (positron) and 820 GeV proton.
The design luminosity is 1.5$\times$10$^{31}$ cm$^{-2}$s$^{-1}$.
Polarized electron and positron beams are available.
Two experimental groups H1 and ZEUS are now engaged in various experiments.
Needless to say, HERA is
expected to be the powerful machine to serve us utmost information on
the nucleon structure.
In addition, HERA has a large potentiality to discover signatures of
the new physics beyond the SM.
This is because of higher energies available
than LEP and cleaner events than hadron colliders such as TEVATRON.
Also the unique signature at the $ep$ collider
which is not obtained from electron-positron colliders nor hadron colliders
reveals us a striking feature of new physics.

The purpose of the present paper is to review the search for sparticles at the
$ep$ collider HERA.
Our emphasis will be paid upon the scalar top quark (stop), the partner of
top quark, characterized by its lighter mass than the top quark and other
scalar quarks (squarks) in a model described later \cite{stop,HK}.
The existence of the light stop could give a clue to the issue on $R_b$
\cite{Rb}.
The expected mass of the stop is clearly within the reach of HERA.
Particularly, the squarks could be singly produced in $ep$ collisions
\cite{stoprb,susana,Hewett}
in the $R$-parity breaking model (RBM) \cite{Barger}
because of the existence of
the electron-quark-squark couplings.
Clearly HERA is the best machine to search for the stop in such models,
because we expect the remarkable peak structure in the Bjorken parameter
$x$ distribution.
We calculate cross sections together with detectable signals
expected to observe at HERA.
Although we have studied production processes for selectron, sneutrino or
photino at HERA \cite{bartl} we do not enter their details because of
their discovery potential to be not so large.

The organization of the paper is as follows.
In Sec.~2 we develop the theoretical framework of
Minimal SUSY Standard Model (MSSM) and the origins of the light stop mass
is discussed.
Decay modes of the stop with mass less than other squarks and gluino are
also exploited.
Experimental status of SUSY particle search is extensively presented in Sec.~3.
Not only status on HERA, but also data on the search from LEP, TEVATRON and
TRISTAN are concisely summarized.
Various analyses on the sparticle production and their expected signatures
from decay processes at HERA are reviewed in Sec.~4.
Stop production by boson-gluon fusion and $R$-parity breaking single
stop production are presented in detail.
The concluding remarks are given in the final section 5.

\section{\it Theoretical framework of MSSM}

\subsection{\it Particle content}
The Minimal SUSY Standard Model (MSSM)  includes the {\it minimal} particle
content.
That is, there should be a new particle (sparticle) for each
known particle in the SM and the two Higgs doublets.
The additional Higgs doublet must be included in order to
produce masses of up- and down-type quarks and to realize the chiral
anomaly cancellation \cite{Nilles}.
The list of particles in the model is shown in
Table {\uppercase\expandafter{\romannumeral 1}} (weak eigenstates).

For quarks and charged leptons there should exist two scalar
partners per species.
These scalars are called squarks and sleptons, generically sfermions.
Before SU(2)$\times$U(1) breaking, the left- and right-sfermions
do not mix since they have different SU(2)$\times$U(1) quantum numbers.
After the breaking, however, they can mix each other.
Actually, this mixing effect is substantial only for the third-generation
sfermions, especially for the stops (superpartners of the top quark).
The left-right mixing of the stops will be discussed  in
the following subsection.

The new fermions are either the superpartner of spin-1 gauge bosons
(gauginos) or that of spin-0 bosons (higgsinos).
They mix each other when SU(2)$\times$U(1) symmetry is broken.
The mass eigenstates (inos) are usually some complicated mixture of
gauginos and
higgsinos.
Electrically neutral-inos and charged-inos are called neutralinos
$\szk$ ($k=1\sim4$)
and charginos $\swi$ ($i=1,2$), respectively.
Gluinos $\gluino$ are free from the mixing since the color SU(3) is not broken.

We need two Higgs doublets in the MSSM \cite{higgs}.
We have five physical Higgs bosons which
remain after SU(2)$\times$U(1) breaking.
They are two CP even ($h^0$, $H^0$)
and one CP odd ($A^0$) neutral scalars and
remaining one charged scalar ($H^\pm$).

\subsection{\it Basic parameters of MSSM}

The standard model has 18 fundamental parameters to be determined
by experiments.
The MSSM has a somewhat larger number of parameters.
They are classified as
({\romannumeral 1}) gauge couplings,
({\romannumeral 2}) superpotential parameters and
({\romannumeral 3}) soft-breaking parameters.

\subsubsection{\it gauge couplings}

The three gauge coupling parameters corresponding to
SU(3), SU(2) and U(1) gauge groups are the same as in the SM.
All gauge interactions are governed by these couplings.
They determine the fermion - sfermion - gaugino interactions
and four-point scalar interactions as well as ordinary
fermion - fermion - gauge-boson interactions.

\subsubsection{\it superpotential parameters and $R$-parity}

In the MSSM ordinary Yukawa interactions are generalized to the
terms in superpotential $W(\hat{\phi})$, where $\hat{\phi}$ denote arbitrary
chiral superfields.
Renormalizability restricts the functional form of the superpotential to
\begin{equation}
W(\hat{\phi})=m_{ij}\hat{\phi}_{i}\hat{\phi}_{j}
+\lambda_{ijk}\hat{\phi}_i\hat{\phi}_j\hat{\phi}_k
\end{equation}
The parameters $m_{ij}$ and $\lambda_{ijk}$ are further constrained by
the gauge symmetry and some discrete symmetries.

A  well-known discrete, multiplicative symmetry is the $R$-parity
defined by
\begin{equation}
R=(-)^{3(B-L)+2S},
\end{equation}
where $B$, $L$ and $S$ denote the baryon number, the lepton number and
the spin, respectively.
This formula implies that all ordinary SM particles
have even $R$-parity, whereas the corresponding superpartners
have odd $R$-parity.
Usually we impose the $B-L$ conservation on the MSSM
and then the MSSM possesses the $R$-parity invariance.
In this case the superpotential has the form ;
\begin{equation}
W=\mu \hat{H}_1 \hat{H}_2 + f_\ell \hat{H}_1 \hat{E^c} \hat{L}
+ f_d \hat{H}_1 \hat{D^c} \hat{Q} + f_u \hat{H}_2 \hat{U^c} \hat{Q},
\label{Rbsp}
\end{equation}
where we have used the superfield notation in
Table {\uppercase\expandafter{\romannumeral 1}}.
Here parameters $f$'s correspond to usual Yukawa couplings and
$\mu$ is a new supersymmetric parameter which has no counterpart in
the SM.

The $R$-parity conservation in scattering and decay processes
has a crucial impact on the SUSY phenomenology at high energy colliders.
First, the SUSY particles must be produced in pairs.
Second, the lightest SUSY particle (LSP) is absolutely stable.
On the other hand, the cosmological constraints tell us that the LSP
should be electrically and also color neutral.
Consequently, it is weakly interacting in ordinary matter.
The heavy unstable SUSY particles must finally decay into the LSP.
Therefore the canonical signature for the $R$-parity conserving
SUSY models at collider experiments
is the large missing (transverse) energy.

It should be emphasized that the $R$-parity conservation is not
automatically imposed in the MSSM.
In fact the $R$-parity breaking superpotential is allowed
by the supersymmetry as well as the gauge symmetry \cite{Barger};
\begin{equation}
W_{\rb}=\lambda_{ijk}\hat{L}_i \hat{L}_j \hat{E^c}_k
+ \lambda'_{ijk}\hat{L}_i \hat{Q}_j \hat{D^c}_k +
\lambda''_{ijk}\hat{U^c}_i \hat{D^c}_j \hat{D^c}_k.
\label{RBW}
\end{equation}
The first two terms violate $L$ and the last term violates $B$.
If we want to explain some unresolved problems such as
({\romannumeral 1}) the cosmic baryon number violation,
({\romannumeral 2}) the origin of the masses and the
magnetic moments of neutrinos and
({\romannumeral 3}) some interesting rare processes
in terms of the $L$ and/or $B$ violation,
the $R$-parity breaking terms must be incorporated in the MSSM.

In such a model the signatures for SUSY particles observed at collider
experiments
should be very different from the caconical one.
We will discuss the typical $\rb$ signals at HERA in Sec.4.

\subsubsection{\it soft-breaking parameters}

SUSY is not an exact symmetry of nature
since our world is not manifestly supersymmetric.
In the MSSM the SUSY breaking is induced by the soft-SUSY breaking
terms, which do not introduce quadratic divergences.
Hence the solution of the naturalness problem remains intact.
There are four types of soft breaking terms ;
({\romannumeral 1})  gaugino masses $M_i$  ($i=1\sim3$),
({\romannumeral 2})  masses for the sfermions $\mtil_{f}$,
({\romannumeral 3})  trilinear term  $A_f$ and
({\romannumeral 4})  three scalar Higgs mass terms.
 These three mass parameters can be re-expressed in terms of
 two Higgs vacuum expectation values, $v_1$ and $v_2$, and
 one physical Higgs mass.
 Here $v_1$ and $v_2$ respectively denote the vacuum expectation values
 of the Higgs field coupled to $d$-type and $u$-type quarks.
 $v_1^2 + v_2^2$ $=$ (246 GeV)$^2$ is determined from
 the experimentally measured $W$-boson mass,
 while the ratio
 \begin{equation}
 \tanbe = {\frac{v_2}{v_1}}
 \end{equation}
 is a free parameter of the model.

 A generally accepted assumption is that
 all the three gaugino mass parameters $M_i$ are equal at some
 grand unification scale $M_X$.
 Then the gaugino mass parameters can be expressed in terms of
 one of them, for instance, $M_2$.
 The other two gaugino mass parameters are given by
 \begin{eqnarray}
 M_1 &=& {\frac{5}{3}}\tan^2{\tew} M_2 \\
 M_3 &=&  {\frac{}{\alpha}}\sw M_2,
 \end{eqnarray}
where $\alpha$ and $\alpha_3$ denote the QED and the QCD coupling constants,
respectively.

\subsection{\it Scalar top in MSSM}
\subsubsection{\it left-right mixing of stops}
In the framework of the MSSM \cite{Nilles},
the stop mass matrix in the ($\st_{L}$, $\st_{R}$) basis is expressed by
\begin{equation}\renewcommand{\arraystretch}{1.3}
{\cal M}^{2}_{\st}=\left(
                 \begin{array}{cc}
                   m^{2}_{\st_{L}} & a_{t}m_{t} \\
                   a_{t}m_{t} & m^{2}_{\st_{R}}
                 \end{array}
                \right),
\label{matrix}
\end{equation}
where $\mt$ is the top mass.
The SUSY mass parameters $m_{\st_{L, R}}$ and $a_{t}$
are parametrized in the following way \cite{susy} :
\begin{eqnarray}
m^{2}_{\st_{L}}&=&{\widetilde{m}}^{2}_{Q_{3}}
  +\mz^{2}\cos{2\beta}\left({\frac{1}{2}}-
{\frac{2}{3}}\sin^{2}\tew\right)+m^{2}_{t}, \label{mstl}\\
m^{2}_{\st_{R}}&=&{\widetilde{m}}^{2}_{U_{3}}
  +{\frac{2}{3}}\mz^{2}\cos{2\beta}\sin^{2}\tew+m^{2}_{t}, \label{mstr}\\
a_{t}&=&A_{t}+\mu\cot\beta.
\end{eqnarray}
The soft breaking masses of the third generation doublet
${\widetilde{m}}_{Q_{3}}$ and the up-type singlet
${\widetilde{m}}_{U_{3}}$ squarks are related to those
of the first (and second) generation squarks as
\begin{eqnarray}
{\widetilde{m}}^{2}_{Q_{3}}&=&{\widetilde{m}}^{2}_{Q_{1}}
-{\widetilde{I}}, \label{mq3}\\
{\widetilde{m}}^{2}_{U_{3}}&=&{\widetilde{m}}^{2}_{U_{1}}
-2{\widetilde{I}}, \label{mu3}
\end{eqnarray}
where ${\widetilde{I}}$ is a function proportional to
the top quark Yukawa
coupling $\alpha_{t}$ and is determined by the
renormalization group equations (RGEs)  \cite{RGE} in the
minimal supergravity GUT (MSGUT).
Throughout this paper we adopt the notation of Ref.\cite{Hikasa}.

   There are two origins for lightness of the stop compared to
the other squarks and sleptons,
{\romannumeral 1}) smallness of the diagonal soft
masses $m^{2}_{\st_{L}}$ and $m^{2}_{\st_{R}}$ and
{\romannumeral 2}) the left-right stop mixing.
Both effects are originated from the large Yukawa interaction of the
top.
The origin {\romannumeral 1}) can easily be seen from
Eqs.(\ref{mstl})$\sim$(\ref{mu3}).
The diagonal mass parameters $m^{2}_{\st_{L}}$ and $m^{2}_{\st_{R}}$
in Eq.(\ref{matrix}) have possibly small values
owing to the negative large
contributions of ${\widetilde{I}}$ proportional to $\alpha_{t}$ in
Eqs.(\ref{mq3}) and (\ref{mu3}).
It should be noted that this contribution is also important in
the radiative SU(2)$\times$U(1) breaking in the MSGUT.
The Higgs mass squared has an expression similar to
Eqs.(\ref{mq3}) and (\ref{mu3}) ;
\begin{equation}
{\widetilde{m}}^{2}_{H_{2}}=
{\widetilde{m}}^{2}_{L_{1}}
-3{\widetilde{I}},
\label{Higgsmass}
\end{equation}
where ${\widetilde{m}}^{2}_{L_{1}}$ denotes
the soft breaking mass of the first generation doublet slepton.
The large contribution of ${\widetilde{I}}$ enables
${\widetilde{m}}^{2}_{H_{2}}$
to become negative at an appropriate weak energy scale.
In order to see another origin {\romannumeral 2}) we should diagonalize
the mass matrix ({\ref{matrix}).
The mass eigenvalues are obtained by
\begin{equation}
m^{2}_{\stl\atop\sth}
         ={\frac{1}{2}}\left[ m^{2}_{\st_{L}}+m^{2}_{\st_{R}}
             \mp \left( (m^{2}_{\st_{L}}-m^{2}_{\st_{R}})^{2}
            +(2a_{t}m_{t})^{2}\right)^{1/2}\right].
\label{stopmass}
\end{equation}
and the corresponding mass eigenstates are expressed by
\begin{equation}
\left({\stl\atop\sth}\right)=
\left(
{\st_{L}\,\cos\tht-\st_{R}\,\sin\tht}
\atop
{\st_{L}\,\sin\tht+\st_{R}\,\cos\tht}
\right),
\end{equation}
where $\tht$ denotes the mixing angle of stops :
\begin{eqnarray}
\tan\tht=
{\frac{a_{t}\,m_{t}}
  {m^{2}_{\st_{R}}-\mstl^{2}}}.
\label{tantht}
\end{eqnarray}
We see that if the SUSY mass parameters and the top
mass are of the same order of magnitude,
small $\mstl$ is possible owing to the cancellation
in the expression (\ref{stopmass}) \cite{stop,HK}.

After the mass diagonalization
we can obtain the interaction Lagrangian for the
mass eigenstate $\stl$.
We note, in particular, that the stop coupling to
the $Z$-boson ($\stl\stls Z$) depends sensitively on
the mixing angle $\tht$.
More specifically, it is proportional to
\begin{equation}
C_{\stl}\equiv {\frac{2}{3}}\sin^{2}\tew - {\frac{1}{2}}\cos^{2}\tht.
\label{c}
\end{equation}
Note that for the special value of $\tht$$\sim$0.98,
the $Z$-boson coupling completely vanishes \cite{DH}.

\subsubsection{R-parity breaking interactions of stop}

Interesting properties of the lighter stop $\stl$
described in the last sub-section
are not modified even in the $R$-parity breaking models.
In this case the stop could have interactions with the ordinary
leptons and/or quraks via the $R$-breaking Lagrangian ;
\begin{equation}
L=\lambda'_{i3j}\cos\tht \stl {\bar{d}}_j P_L \ell_{i} + h.c.,
\end{equation}
which is originated from the superpotential (\ref{RBW}).
In particular the interaction with the electron ;
\begin{equation}
L=\lambda'_{131}\cos\tht (\stl {\bar{d}} P_L e + \stls \bar{e} P_R d)
\label{stRb}
\end{equation}
will be most suitable for the $ep$ collider experiments at HERA
because the stop will be produced in the $s$-channel in $e$-$q$ sub-processes
\cite{stoprb}.
Note that the stop can not couple to any neutrinos via $R$-breaking
interactions.
This is a unique property of the stop which could be useful for us to
distinguish
the stop from some leptoquarks.

\subsubsection{decay modes of stop}

Here we examine the decay modes of the stop.
In the MSSM, the stop
lighter than the other squarks and gluino
can decay into the various final states :
\begin{eqnarray*}
\stl &\to& t\,\szk   \qquad\qquad\qquad\qquad\qquad\qquad({\rm a}) \\
 &\to& b\,\swi   \ \ \quad\qquad\qquad\qquad\qquad\qquad({\rm b})\\
 &\to& b\,\ell\,\sneu \qquad\qquad\qquad\qquad\qquad\qquad({\rm c})\\
 &\to& b\,\nu\,\slepton \qquad\qquad\qquad\qquad\qquad\qquad({\rm d})\\
 &\to& b\,W\,\szk \ \qquad\qquad\qquad\qquad\qquad \ ({\rm e})\\
 &\to& b\,f\,\overline{f}\,\szk \qquad\qquad\qquad\qquad\qquad
 \ ({\rm f})\\
 &\to& c\,\sz1, \quad\qquad\qquad\qquad\qquad\qquad \ ({\rm g})\\
 &\to& e\,d, \ \ \quad\qquad\qquad\qquad\qquad\qquad \ ({\rm h})
\end{eqnarray*}
where $\szk$ ($k=1\sim 4$), $\swi$($i=1,2$), $\sneu$ and $\slepton$,
respectively, denote
the neutralino, the chargino, the sneutrino and the charged slepton.
(a) $\sim$ (g) are the $R$-parity conserving decay modes, while (h) is realized
by the RB couplings (\ref{stRb}).

If we consider the stop with mass small enough
in the $R$ conserving case,
the first five decay modes (a) to (e) are kinematically
forbidden due to the observed top mass $m_{t}$$\simeq$175 GeV \cite{top}
as well as the model independent
lower mass bounds for sparticles ;
$m_{\swl}$ $\nge$ 45 GeV, $m_{\slepton}$ $\nge$ 45 GeV and
$m_{\sneu}$ $\nge$ 40 GeV.
So (f) and (g) survive.
Hikasa and Kobayashi \cite{HK} have shown that
the one-loop mode $\stl\to c\sz1$ (g) dominates over the
four-body mode $\stl\to bff'\sz1$ (f).
So we can conclude that such a light stop will decay into
the charm quark jet plus the missing momentum taken away
by the neutralino with  almost 100$\%$ branching ratio.
On the other hand,
if we consider the RB coupling $\lambda'_{131}$ $>$ $0.01$,
which roughly corresponds to the coupling strength detectable at HERA,
the decay modes (c) to (g) are negligible due to their large power of
$\alpha$ arising from
multiparticle final state or one loop contribution.
So there the two body modes (a), (b) and (h) survive in this case.


\section{\it  Experimantal status of SUSY particle search}

   Most of the theoretical models assume $R$-parity conservation, which has
important physical implication: (1)  SUSY particles must be produced in pairs,
(2) heavy SUSY particles decay to lighter SUSY particles, and (3)  the LSP is
stable.   The phenomenology of  $R$-parity violating interactions differs from
that of the MSSM  in two main aspects \cite{Dreiner}:
\begin{enumerate}
\item    the LSP is no longer stable since it is not protected by  symmetry,
that is, it can decay within the detector; and
\item   it is possible to have single production of supersymmetric particles,
since the final state is no longer restricted to be $R$-parity even.
\end{enumerate}
We present the experimental status of SUSY particle search in $R$-parity
conserving case and
$R$-parity violating case separately.

\subsection{\it   $R$-parity conserving case}
\subsubsection{\it   Sneutrinos and sleptons}
    Once enough energy is available, the most efficient way to produce
superpartners is to produce them in   pair in high-energy leptonic or hadronic
interactions.     Sneutrinos, which are most likely invisible like neutrinos,
can be  produced  in pair in the $Z^0$  boson decay at  LEP.
The partial decay width for  sneutrinos  is related to that for neutrinos by
\cite {Fayet}
\begin{eqnarray}
\Gamma  (Z^0 \rightarrow  {\widetilde{\nu}}_\ell
\bar{{\widetilde{\nu}}_\ell} ) =
            \frac {1}{2} \beta^3 \Gamma (Z^0 \rightarrow  \nu_ \ell
\bar{\nu}_\ell),  \qquad
              \ell= e, \mu  ~~ {\rm or}~~    \tau,  \nonumber
\end{eqnarray}
where  $\beta = v_{\widetilde{\nu}}/c$.   The invisible decay width of the
$Z^0$ measured at LEP is in good agreement with the standard model  of
$N_\nu$=3 within errors.   Therefore any additional contribution to  the
invisible decay width of  the  $Z^0$   is limited by
\begin{eqnarray}
\Delta  \Gamma{{invis.}\atop Z} < 8.8 {\rm MeV},  \nonumber
\end{eqnarray}
at the 95\% CL.  This  leads  to the mass limit  of  $m_{\widetilde{\nu}}$ $>$
43 GeV  if the three sneutrinos are degenerate in mass, while
$m_{\widetilde{\nu}}$ $>$ 40 GeV  for  one generation  sneutrino
\cite{Komamiya}.

 Events of the charged slepton pair production through  the $Z^0$
decay
\begin{eqnarray}
 Z^0 \rightarrow {\widetilde{\ell}}^+ {\widetilde{\ell}}^-
\rightarrow  \ell^+ \ell^-   + 2  \
  {unobserved \ neutralinos}  \nonumber
\end{eqnarray}
are characterized by  non-coplanar  lepton pairs with large missing
energy -momentum  in the detector.
The expected main backgrounds are
 $Z^0 \rightarrow \tau^+ \tau^- (\gamma),  Z^0  \rightarrow e^+ e^-\ell^+
\ell^- ,   Z^0  \rightarrow e^+ e^- +  hadrons$. The 95\% CL lower mass
limit on   $\widetilde {e}, \widetilde{\mu}$   and $ \widetilde{\tau}$
obtained at LEP is  about  45 GeV almost independent of slepton
species\cite{Komamiya}.

\subsubsection{\it   Gluinos and squarks}
   Since gluinos and squarks are strongly interacting,
they would be the SUSY particles with the largest cross sections at
 $\bar {p} p$ colliders.   The branching ratios of gluinos and squarks
decaying into  various chargino and neutralino states depend on
 respective masses and mixing angles.
In a scenario of a low mass gluino and massless photino
($\widetilde{\gamma}$ ) as the LSP,  the decay modes become very simple.
If $m_{\widetilde {g}} < m_{\widetilde {q}}$ , the squark decays dominantly
into $\widetilde {g} \bar {q} $ and the main gluino decay is $\widetilde {g}
\rightarrow  q\bar {q}  \widetilde{\gamma}$.   If    $m_{\widetilde {q}}
< m_{\widetilde {g}}$ ,  then the gluino decays dominantly into
$\widetilde {q} \bar {q}$ and the main squark decay is  $\widetilde{q}
\rightarrow   q\widetilde{\gamma}$.   Since the photinos escape the
detectors without any signature, the SUSY events would have two or
more jets with a large amount of imbalanced transverse momenta.

 The limits on  $m_{\widetilde {q}}$  and  $m_{\widetilde {g}}$,
presented by the CDF experiment, are based on the comparison of the
observed missing transverse energy (${\ooalign{\hfil/\hfil\crcr$E$}}_T$ )
distribution with predictions for the standard model background and the QCD
background plus SUSY contribution estimated by the ISAJET Monte Carlo
samples\cite{Hu}. Squark and gluino mass limits obtained at the CDF
 experiment , under the condition of SUSY parameters $\mu = -250$ GeV,
$\tan\beta$ = 2 and $m_H$=500 GeV, are shown in Fig.1 together with those
of D0, LEP and UA1/UA2
experiments\cite{Hagopian}.
  A search for squarks and gluinos was made by
D0 in the three or
more jets plus ${\ooalign{\hfil/\hfil\crcr$E$}}_T$  channel.
The number of events observed in the data sample from an
integrated luminosity of 13.4 $\pm$1.6 pb$^{-1}$ was consistent
with background.   The limits on  $m_{\widetilde {q}}$  and
$m_{\widetilde {g}}$ plane  are shown in Fig.1 .
For heavy  squarks, a lower  gluino mass limit of 146 GeV
was obtained, and for equal squark and gluino masses a mass limit
of 205 GeV was obtained at the 95\% CL \cite{Hagopian}.

\subsubsection{\it   Charginos and neutralinos}
Charginos  are the mass eigenstates corresponding to the linear
combination of winos and charged Higgsinos.
Charginos have been searched for at LEP via their pair production and decays
\begin{eqnarray*}
         e^+ e^-   &\rightarrow&   Z^0  \rightarrow
{\widetilde {W}_1}^+   {\widetilde {W}_1}^-     ,  \\
         {\widetilde {W}_1}^+
 &\rightarrow&
q \bar {q} \widetilde {Z}_1 ,   \ell \nu \widetilde {Z}_1,  \\
         {\widetilde {W}_1}^+
  &\rightarrow&
 \ell^+ \widetilde {\nu}_\ell .
\end{eqnarray*}
The mass limit at L3 comes from the line-shape measurement constraint :
$m_{\widetilde {W}_1}$ $>$ 44 GeV.
This result is independent from the
chargeno decays and the field contents(wino and charged Higgsino mixing)
\cite{Zhou}.
   Searches for the signatures of acoplanar leptons, acoplanar jets and
isolated particles were combined at ALEPH.   Irrespective of the field
content and of the mass of the LSP, masses for chargeno below 45.2 GeV
are excluded at the 95\% CL\cite{Medcalf}.

Neutralinos, denoted by $\widetilde {Z}_1 , \widetilde {Z}_2,\widetilde {Z}_3 ,
\widetilde {Z}_4$  in order of increasing mass, are linear combinations of
the photino, the zino, and the Higgsinos.   The neutralino pair production
at LEP is searched for via the following processes:
\begin{eqnarray}
  e^+ e^-  \rightarrow  Z^0 \rightarrow  \widetilde {Z}_1
\widetilde {Z}_1,  \widetilde {Z}_1  \widetilde {Z}_2 ,
 \widetilde {Z}_2  \widetilde {Z}_2.   \nonumber
\end{eqnarray}
In  most cases, the main $\widetilde {Z}_2$   decay mechanism is
\begin{eqnarray}
  \widetilde {Z}_2   \rightarrow  \widetilde {Z}_1 Z^*
\rightarrow  \widetilde {Z}_1  f \bar {f}  \ {\rm or} \  \widetilde {Z}_2
\rightarrow  \widetilde {Z}_1  \gamma. \nonumber
\end{eqnarray}
Since the  $\widetilde {Z}_1$  escapes undetected, the signature of the
neutralino pair production would be excess in $\Gamma{{invis.}\atop Z}$
measurement. The signature of the above $\widetilde {Z}_2$
decay processes are missing energy due to the undetected
$\widetilde {Z}_1$ and one or two photons, two or four acollinear
and acoplanar leptons, or one to four hadronic jets from the primary
quarks \cite{Acciarri}.

From the absence of any candidate of   $\widetilde {Z}_1$  and
$\widetilde {Z}_2$    at DELPHI\cite{Lutz} and L3\cite{Acciarri},
limits on the probability that the $ Z^0$  would decay into these
channels were set.  Combining these results with those  from the $Z^0$
widths  within  the MSSM, one can caluculate, for every combination of $M,
\mu$ and $\tan\beta$ values, the total contribution  of $ Z^0$ decays  into
neutralinos and charginos, and can deduce  the excluded regions in the $(M,
\mu)$
 plane.     The  excluded regions for  four values of  $\tan\beta$, obtained by
the L3 experiment through search for
$e^+ e^-  \rightarrow  Z^0 \rightarrow  \widetilde {Z}_1  \widetilde {Z}_2$
or  $\widetilde {Z}_2  \widetilde {Z}_2 $ are shown in Fig.2
\cite{Acciarri}.
For moderate or high values of  $\tan\beta$,  a significant part of the
accessible parameter space is excluded.
All neutralino masses are functions of the parameters $M$, $\mu$ and
$\tan\beta$.   Therefore, constraints on the MSSM parameter space translate
into limits on
these masses which  are shown in Fig.3 as a function of
$\tan\beta$\cite{Acciarri}.

   Data  taken at the TOPAZ and VENUS  with  the TRISTAN $e^+ e^-$ collider
have been analyzed to study single photon events.  The TOPAZ , based on
data of an integrated luminosity of 164 pb$^{-1}$  at  the energy $\sqrt{s}=$
 59 GeV,  observed 4 single photon candidates remained after event selection,
which are consistent with the prediction of the standard model  plus
background \cite{Abe}.

The VENUS , based on  data of  225 pb$^{-1}$  at  $\sqrt{s} = $ 58 GeV,
measured the single photon cross section , which is consistent with
expectation.  No anomalous signal has been observed leading to a lower
limit on the mass of SUSY particles under assumption of radiative pair
production of photinos.  For massless photinos, the scalar electrons
of the mass degenerate case are excluded below  51.9 GeV at
the 90\% CL  \cite{Hosoda}.   Combining this    value with     other
single photon experiments, the lower  limit on the scalar electron
was determined to be 72.6 GeV at the 95\% CL.
 Fig.4 shows the excluded region   on   the ($m_{\widetilde {\gamma}},
m_{\widetilde {e}}$)  plane together with the results from other
 experiments  compiled  in Ref.\cite{Hosoda}.

Bounds on the gaugino  parameters $\mu , \tan\beta$ and $M_2$  have been
calculated from experimental data\cite{Kon}.   Fig.5 shows the region in
the ($\mu , M_2$ ) plane  for $\tan \beta =2$ excluded by the experimental
data  on
\begin{description}
\item[A] lower bound on the mass of lighter chargino
$m_{\widetilde {W}_1} >$ 45 GeV,
\item[B]upper bound on the branching ratio of the visible
neutralino mode $Br(Z^0\rightarrow vis.) < 5\times10^{-6}$,
\item[C] upper bound on the invisible width of the $Z^0$  , and
\item[D] accepted gluino mass bound at CDF $m_{\widetilde {g}} >$150 GeV
(90\% CL),
\end{description}
where the hatched regions  of each contour have been excluded.
We see that the stringent bound comes from the constraint on the
visible width of the $Z^0$ at LEP (B), or from the gluino search at CDF(D).

\subsubsection{\it    Stop}
 Among squarks, it may be that the stop is very light because of the
high mass of the top. For  a particular value of the mixing angle
between ${\widetilde {t}_R}$   and ${\widetilde {t}_L}$ , the lighter mass
eigenstate ${\widetilde {t}_1}$  decouples from the $ Z^0$ .
Pair production of light stop has been extensively searched for by  OPAL
in   $e^+ e^-$ collisions at LEP \cite{Akers}.   The decay mode
$\widetilde {t}_1  \rightarrow c\widetilde {Z}_1$ was searched for
in the data of the integrated luminosity of  69.1 pb$^{-1}$  ,
which corresponds to $1.68\times10^6$ produced $Z^0 \rightarrow
\bar {q}q$ events.   With no ${\widetilde {t}_1}$ candidates,
the region where more than 3.0 events are expected is excluded at 95\% CL.
Fig.6 shows the excluded region  in the ($\theta_t , m_{\widetilde {t}_1} $)
plane for the case of the mass difference $\Delta m$ between
${\widetilde {t}_1}$ and ${\widetilde {Z}_1}$  larger than 5 GeV
together with the region excluded by lower energy experiments\cite{Akers}.
The structure  in the  OPAL limits  on  $\theta_t $ in Fig.6 is due to
decoupling of  the ${\widetilde {t}_1}$ from the $Z^0$  for $\theta_t
\sim0.98$.
The OPAL  excludes the existence of the  ${\widetilde {t}_1}$  with a mass
below
45.1 GeV at the 95\% CL, where $\theta_t $ of  ${\widetilde {t}_R}$
and ${\widetilde {t}_L}$  is smaller than 0.85 or greater than 1.15,
and $\Delta m$ is greater than  5 GeV.
The exclusion regions
in the ($m_{\widetilde {t}_1},  m_{\widetilde {Z}_1}$) plane are shown
in Fig.7 for various $\theta_t$.
The DELPHI  has excluded significant regions in the
($m_{\widetilde {t}_1} , m_{\widetilde {Z}_1}$) plane
for different mixing angle  $\theta_t$
by  assuming  that $\widetilde {t}_1  \rightarrow  c \widetilde {Z}_1$,
the most likely decay mode allowed by existing limits\cite{Hultqvist}.
The stop below 46 GeV has been excluded for any value of the mixing
angle,  except for a small window in
$\theta_t \sim 1$ \cite{Lutz}.

A search for  the light stop has been carried out by  the VENUS  at
TRISTAN\cite{Shirai}.  A data sample of 210  pb$^{-1}$  has been
analyzed to find events with large acoplanar particle groups, assuming
the two-body decay  $\widetilde {t}_1  \rightarrow  c \widetilde {Z}_1$.
The observed number of events was consistent with that expected from the
known processes.  They obtained the mass limits in  the ($m_{\widetilde
{t}_1} ,
m_{\widetilde {Z}_1}$) plane at  95\% CL, where ${\widetilde {t}_1}$
was excluded for the mass region  from 7.6 GeV to 28.0 GeV
for massless  LSP .
   The  inclusive $D^{*\pm}$  production cross section in two-photon processes
was measured with the TOPAZ detector\cite{EnomotoPR}.  The differential cross
sections $d \sigma(D^{*\pm})/dp_T$  obtained were compared with theoretical
predictions, such as those involving direct and resolved photon processes.
A discussion  was made on  interpretation of
the data of $d \sigma(D^{*\pm})/dp_T$  assuming stop pair production
decaying into  a charm quark and a neutralino\cite{Enomoto}.

   A constraint from $ b \rightarrow s \gamma$ process to the
MSSM has been derived in the light stop region \cite{Okada}.
It was  pointed out that although some region in the parameter
space is excluded from this process there remains a large parameter
space where the amplitude of the  $ b \rightarrow s \gamma$ is
suppressed due to partial cancellation between  different diagrams.
Stops as light as 20 GeV are  still viable from the  $ b \rightarrow s
\gamma$  constrainrt.   It is also pointed out by Fukugita et al.
that the light stop with a mass  as small as  $m_{\widetilde {t}}
\sim$ 20 GeV is still an allowed possibility within the standard,
minimal SUSY GUTs with the SUSY breaking induced by minimal
supergravity \cite{Fukugita}.

Motivated by the fact that the stop may be considerably lighter than other
squarks,  H. Baer, J. Sender and X. Tata have reinvestigated its signals at
the Fermilab Tevatron with simulations using ISAJET 7.07 under the assumption
that it decays via  $\widetilde {t}_1 \rightarrow  b\widetilde {W}_1$ or
$\widetilde {t}_1  \rightarrow  c \widetilde{Z}_1 $.  They have shown that
experiments should be able to probe stop  mass up to $\sim 100$ GeV  with
an integrated luminosity of 100 pb$^{-1}$  \cite{Baer}.

\subsection{\it   $R$-parity violating case}
\subsubsection{\it     Constraints from low-energy processes and neutrino
physics }

   We are particularly  interested in the case in which the term
$\lambda '_{1jk}\hat{L}_1\hat{Q}_j{\hat{D}_k}^c$  in Eq. (4) in Sec.2
is dominant, since this leads to resonant squark production through the
$s$-channel in the $e$-$q$ subprocess at HERA \cite{Butterworth}.

The term  $\lambda '_{11k}\hat{L}_1\hat{Q}_1{\hat{D}_k}^c, (k=1,2,3)$
contributes to the semileptonic decays of quarks, and the term
$\lambda '_{12k}\hat{L}_1\hat{Q}_2{\hat{D}_k}^c, (k=1,2,3)$
contributes  to forward-backward asymmetries measured in  $ e^+e^-$
collisions\cite{Barger}.  Existing limits for these coupling constants
are given in \cite{Barger}as a function of the mass of squarks.
The term $\lambda '_{1j1}\hat{L}_1\hat{Q}_j{\hat{D}_1}^c, (j=1,2,3)$
also contributes to atomic parity violation\cite{Barger}.
The   $1\sigma$ bound for the coupling of this term is
\begin{eqnarray}
 \lambda '_{1j1}  < 0.26 (m_{\widetilde {q}
{j \atop L}} /100 {\rm GeV}),  \nonumber
\end{eqnarray}
where the effect of the radiative corrections is taking into
account\cite{Barger}.
   A general analysis of the constraints derived from neutrino
physics on explicit $R$-parity breaking in supersymmetric models
is presented\cite{Enqvist}.    The upper bounds for $\lambda _{ijk}$
and  $\lambda '_{ijk}$  were obtained through  (i)neutrino oscillations,
(ii) neutrinoless double beta decay and (iii) neutrino cosmology and
astrophysics, which were stringent in particular for lepton number
violation involving the third generation\cite{Enqvist}.

\subsubsection{\it    HERA  }
$R$-parity violating SUSY particles  as well as leptoquarks,
leptogluons and excited leptons  are being  searched for at
the HERA $ep$ collider  by  H1.  In a data sample of
$\approx$ 0.5 pb$^{-1}$ no evidence was found for any squark
production in the  mass between 45 GeV up to 275 GeV
\cite{Abt,Kohler}.  In the supersymmetric model,
the single production of squark is allowed  through $R$-parity
violating couplings.  If the model is maximally $R$-parity
violating or if the photino mass is larger than the squark mass,
these states are indistinguishable from leptoquarks.
It implies the rejection limits for a search of squarks from
$R$-parity violating supersymmetry.   The rejection limits on
the coupling $\lambda '_{111}$   are shown in Fig.8 as a function
of the mass of squark   for various fixed photino masses \cite{Kohler}.

   The new particles are being searched for at HERA by ZEUS.
No evidence for leptoquark, leptogluon,  squark, or excited
electron production was found in a data sample of
26 nb$^{-1}$ \cite{McLean}.
Limits on the production of squarks were determined for
masses above 25 GeV.  Using the leptoquark limits,
one can derive limits on the $R$-parity violating coupling at the 98\% CL
as shown in Fig.9.  For electroweak coupling ($\sqrt{4\pi/\alpha}$),
this yields lower limits on squark masses at the 95\% CL :
$m_{\widetilde {d}} >$168 GeV and $m_{\widetilde {t},\widetilde {u}} >$
92 GeV \cite{McLean}.

\section{\it Production processes at HERA}

\subsection{\it Scalar electron production processes}

HERA is the world first $ep$ collider, at which we can observe
various inelastic processes,
i.e., $eq$, $eg$, $\gamma g$ collisions and so on,  in addition to the elastic
processes $ep\to Xp$.
For the purpose of the sparticle search, each collision processes could be
useful for us to detect signatures from different kind of sparticles.
In this subsection we discuss the scalar electron (selectron) production
at HERA.

\subsubsection{\it R-parity conserving case}

The simplest sparticle production process at HERA would be the SUSY
neutral current process \cite{susync,ncour}
\begin{equation}
ep\to \sele\sqk X,
\label{SNC}
\end{equation}
where $\sele$ and $\sqk$ denote the selectron and squark, respectively.
Even in such a simple process, $\sele$ and $\sqk$ could be produces at
the same time and we could get information of both masses of $\sele$ and $\sqk$
from analyses of the process.
This is a remarkable advantage of the $ep$ collider HERA.
Phenomenological analyses for the process have been given by many authors
\cite{bartl,ncex}
and we can find that the total cross section of the process will be
$\sigma \nge $0.1 pb for $m_{\sele} + m_{\sqk}$ $\nle$ 150 GeV.
Unfortunately, however, the TEVATRON experiment has been excluded a light
squark, $m_{\sqk} \nle$ 150 GeV, by the negative search for the squark pair
production $p\bar{p}\to \sqk\sqk^* X$ \cite{Hu,Hagopian},
where they assume 5 flavor squarks
are degenerated in mass.

Therefore, we should say that the search for $\sele$ and $\sqk$ by (\ref{SNC})
will be not so hopefull.

{}From the same reason the SUSY charged current process \cite{susync,ccour}
\begin{equation}
ep\to \sneu\sqk X
\label{SCC}
\end{equation}
and the squark pair production via boson-gluon fusion \cite{bg}
\begin{equation}
ep\to e\sqk\sqk^* X  \qquad \qquad (\sqk \neq \st)
\label{BGSQ}
\end{equation}
will have too small cross section for us to extract from some
large background processes.
Note that the large lower mass bound from the TEVATRON could not be
applicable to the stop.

The SUSY bremsstrahlung process
\begin{equation}
ep\to \sele\sz1 q X
\label{SBRMS}
\end{equation}
has also been analyzed by several authors \cite{susybr,tsutsui,ncour}
since this process has the dominant contribution from
the $t$-channel photon exchange diagram which is not affected by
the existence of heavy squark.
However, the total cross section for this process will be rather small
$\sigma \nge $0.01 pb even for the light selectron, $m_{\sele} \nle$ 50 GeV,
which has been almost excluded by the LEP.

Up to now we have discussed the inelastic processes only.
On the other hand, the elastic selectron production
\begin{equation}
ep\to \sele\sz1 p
\label{ELSE}
\end{equation}
has been known to have rather large and viable cross section
\cite{tsutsui,elastic}.
In particular, Lopez et al. \cite{lopez} have pointed out recently that
the Leading Proton Spectrometer at HERA would be efficient detector,
which can measure momentum of the leading proton scattered
in very forward region.

 The slepton - squark production in the NC and CC processes at
 LEP $\otimes$ LHC whose total available energies are larger than HERA
 is also discussed in Refs.\cite{lhc,thomas}.

\subsubsection{\it R-parity violating case}

If we consider the $R$-parity breaking moel,
the experimental signatures of sparticles will be
completely different from those in the $R$-parity
conserving case as has been shown.
First, the lightest sparticle will not be stable and will decay into
the ordinary particles via the R-parity interaction.
Second, some sparticles could be singly produced in $eq$ sub-processes
at HERA via the R-parity coupling $\lambda'$ in Eq.(\ref{RBW}).

In the former aspect, Dreiner et al. \cite{rbel}
have shown that the SUSY neutral current
process Eq.(\ref{SNC}) will have remarkable signature and
the mass reach at HERA will become about
$m_{\sele} + m_{\sqk}$ $\nle$ 200 GeV even for very small R-breaking
couplings $\lambda'\nge$ $10^{-6}$.

The extensive study for the latter situation has been given by
Butterworth et al. \cite{susana} and by ourself \cite{stoprb,sthera,stopsg}.
The single production of the SUSY partnars of light quarks has been
discussed in Refs.\cite{susana,Hewett}.
On the other hand, we have published some papers in which considered
the single stop production at HERA.
Our works will be summarized in next subsection.
Butterworth et al. have analyzed extensively
the single production of squarks with RB couplings
in the first and second generation \cite{susana}.
For example,
\begin{equation}
e^- u \to \sd
\end{equation}
could be possible and the $d$-squark will decay via $R$-conserving couplings
\begin{equation}
\sd \to d \sz1.
\end {equation}
Moreover, the lightest neutralino $\sz1$ will decay into
$u\bar{d}e^-$ or $\bar{u}de^+$ through
the $R$-breaking interaction proportional to $\lambda'$.
Here the interesting point is that the decay product could contain not only
the electron but also the positron.
This is because the neutralino is the Majorana fermion.
In fact, the positron signature from the electron-proton collision
will be distinguishable from almost standard background.
In their model of cascade decays of the squarks to the LSP,
the squark mass reach was formed to be at HERA
\begin{eqnarray}
       m_{\widetilde{q}}\stackrel{<}{\sim} 270 {\rm GeV} \ {\rm for} \
\lambda '
       \stackrel{>}{\sim}0.08
\nonumber
\end{eqnarray}
 and the reach in the Yukawa coupling is
\begin{eqnarray}
       \lambda ' \stackrel{>}{\sim}5.3{{\times}}  10^{-3} \  {\rm for} \
       m_{\widetilde{q}}
       \simeq 100 {\rm GeV}.
\nonumber
\end{eqnarray}

\subsection{\it  Stop production processes}
\subsubsection{\it  Boson-gluon fusion}
The purpose of the present subsection is to discuss  the production of the
light stop at HERA in the framework of the MSSM with the $R$-parity
conservation.   The possible existence of the light stop  $\widetilde{t}_1
$ has been
discussed by some people
\cite{stop,HK,DH,Kon,Fukugita}.
The most promissing production process in $ep$ collisions
will be the boson-gluon fusion(BGF)\cite{Schuler,bg,stopbg} :
\begin{equation}
      e^\pm p     \rightarrow      e^\pm    \widetilde{t}_1
\widetilde{t}^{*}_1   X .
\end{equation}
Its Feynman diagrams are depicted in Fig.10.
Although it will be shown that the Weizs\"{a}cker - Williams approximation
(WWA) is very appropriate to our purpose, we first carry out the exact
calculations
\begin{equation}
          d\sigma  =   \frac{1}{2 \eta s}   G(\eta,\hat{s})   \int d\eta
dPS^{(3)}
              {\times}  \frac{1}{8}   \sum_{spin}   \bigl | M_a + M_b + M_c
\bigr | ^2 .
\end{equation}
Here $\eta$ denotes the momentum fraction of gluon in the proton and
$\hat{s}  \equiv  (p_f + p_{f'})^2$.  The gluon distribution function
$G(\eta,\hat{s})$
is given by the set 1 of Ref.\cite{Eichten} and the three-body phase space
volume
is expressed as
\begin{equation}
dPS^{(3)}=\frac{1}{128 \pi^3}\Theta (s+2m^{2}_e -2\sqrt{m^{2}_e (s+m^{2}_e)}
-W^{2}_1)
   dy dQ^2 dz  \frac{d\Phi}{2\pi}
\end{equation}
with
\begin{eqnarray}
&&W^{2}_1 \equiv (2m_{\widetilde{t}_1} + m_p)^2-m ^{2}_p,
\nonumber\\
&& Z    \equiv p \cdot p_f/p\cdot q,
              \nonumber\\
&&\cos\Phi \equiv \frac{  ({\bf  p}{\times} {\bf    l}_e )\cdot
({\bf p}{\times} {\bf p}_f )  }
{   |{\bf  p}{\times} {\bf    l}_e |\cdot  | {\bf p}{\times} {\bf p}_f |  }.
\nonumber
\end{eqnarray}
The invariant amplitudes corresponding to Fig.10 are, respectively, given by
\begin{eqnarray}
    && M_a  = \frac{i e^2 g_s }{Q^2} \frac{(q-2p_{f'})_\mu (p-2p_f)_\nu}
                      {m^{2}_{\widetilde{t}_1} - \hat{t}} T^{\mu \nu},   \\
    && M_b  = \frac{i e^2 g_s }{Q^2} \frac{(q-2p_{f})_\mu (p-2p_f')_\nu}
                                 {m^{2}_{\widetilde{t}_1} - \hat{u}} T^{\mu
\nu},   \\
    && M_c  =- \frac{2i e^2 g_s }{Q^2} g_{\mu \nu} T^{\mu \nu}
\end{eqnarray}
with  $\hat{t}=(p_f - p)^2 , \hat{u}=(p_{f'} - p)^2$ and
\begin{eqnarray}
&&T^{\mu\nu}  =  \Big[\bar{u}(l)\gamma^\mu
\Bigl(\frac{2}{3} + C_{{\widetilde{t}_1}}
 \frac{Q^2}{Q^2+m^{ 2}_Z}(A^{e}_L  P_L +A^{e}_R  P_R)\Bigr)u(l_e) \Big]
\epsilon_\nu (p),
   \nonumber\\
&&C_{\widetilde{t}_1}
=\frac{\cos^2\theta_t-\frac{4}{3}\sin^2\theta_W}{2\cos\theta_w\sin\theta_W},
        \nonumber\\
&&A^{e}_L = \frac{1}{2}(\cot\theta_W-\tan\theta_W),           \nonumber\\
&&A^{e}_R  =  -\tan\theta_W,
\nonumber\\
&&P_{L \atop R }= \frac{1}{2}(1  \mp  \gamma_5).
\nonumber
\end{eqnarray}
It is worth mentioning that the constant $C_{\st_{1}}$
representing the strength of $\st_{1}^{*}\st_{1}Z$
coupling depends sensitively on the mixing angle $\tht$ \cite{DH}.
In particular, $C_{\st_{1}}=0$ if
$\tht =\cos^{-1}({{2}\over{\sqrt{3}}}\sin{\tew})\simeq 0.98$.
On the other hand, $\st_{1}^{*}\st_{1}\gamma$ and
$\st_{1}^{*}\st_{1}g$ couplings do not depend on $\tht$.
Therefore, the $Z$ boson contribution to the cross section
depends on $\tht$, while that of the photon is independent from $\tht$.

  It is interesting to compare our exact tree level
calculation with the
Weizs\"acker-Williams approximation (WWA) ;
\begin{equation}
{{\ddf\sigma}\over{\ddf z}}=
\int \ddf yP(y)\int \ddf\eta G(\eta ,\hat{s}')
{{\ddf\hat{\sigma}}\over{\ddf z}},
\end{equation}
where $\hat{s}'\equiv y \eta s$.
The WWA factorizes the cross section of the process, shown in
Fig.~10, into the probability for emitting photon from lepton ;
\begin{equation}
P(y)={{\alpha}\over{2\pi}}
{{1+(1-y)^{2}}\over{y}}\log{{{Q^{2}_{max}}\over{Q^{2}_{min}}}}
\end{equation}
and the BGF cross section involving real photon ;
\begin{equation}
{{\ddf\hat{\sigma}}\over{\ddf z}}=
{4\over9}\pi\alpha\alpha_{s}
{{1}\over{\hat{s}'^{3}z^{2}(1-z)^{2}}}
[2m_{\st_{1}}^{4}-2m_{\st_{1}}^{2}\hat{s}'z(1-z)
+\hat{s}'^{2}z^{2}(1-z)^{2}].
\end{equation}

  The numerical calculations have been performed by using the
program packages BASES \cite{bases}. The standard model parameters are
taken as $\alpha =1/137$,  $\sin^{2}\tew =0.23$ and $m_{Z}=91.1$ GeV.
In Fig.~11 we show the stop mass dependence of the total
cross sections $\sigma$ in the exact tree level calculation and
in the WWA. It is found that the $\gamma$ contribution is much
larger than the $Z$ boson contribution, therefore the WWA is a
good approximation. As mentioned already the photon
contribution does not depend on the stop mixing angle $\tht$.
The total cross section, consequently, is insensitive to $\tht$.
  Throughout the calculations in Fig.~11,
we have adopted the
lower cut for $Q^2$ as $Q^2 >5$ GeV$^{2}$
in order to make detectable the scattered electron.
Because of the photon
dominance, the total cross section significantly increases when smaller
$Q^2$ cuts are adopted.
If $Q^2$ is cut at the kinematical limit ;
$Q^{2}>m_{e}^{2}{{y^{2}}\over{1-y}}$, the total cross section
is about four times as large as that with cut $Q^{2}>5$ GeV$^{2}$.
If we adopt $Q^2$ cut
$Q^{2}>m_{e}^{2}{{y^{2}}\over{1-y}}$,
the detectable cross section turns out to be
$\sigma {{\mathstrut >}\atop{\sim}}
0.1$pb in the case of
$\mstl {{\mathstrut <}\atop{\sim}}50$ GeV.

\subsubsection{\it  Resonance production}

   In the present subsection, we shall concentrate our discussions to
   $R$-parity breaking (RB) process. We start from a coupling of the stop
   $\widetilde{t}_1 $
\begin{eqnarray}
{\cal L}_{int} = {\lambda'_{131}} \cos\theta_t(\widetilde{t}_1 \bar {d}
P_L e + \widetilde{t} ^{*}_1 \bar{e} P_R d )
\end{eqnarray}
originated from the RB superpotential Eq.(\ref{RBW}).
Here $P_{L,R}$ denore left and right  handed chiral projection operators.
The coupling Eq. (\ref{stRb}) is the most suitable for the $ep$ collider
experiments at HERA, because the stop will be produced through the $s$-channel
in the e-d subprocess as shown in Fig.~12 for example,
\begin{eqnarray}
       e^\mp p \rightarrow   (\widetilde{t}_1  X )    \rightarrow  e^\mp q X.
\end{eqnarray}
  For simplicity, we will assume $\lambda'_{131} $ to be only non-zero
  coupling parameter in the following.   The upper bound on the strength of
  coupling  has been investigated through the low-energy experiments
  \cite{Barger} and
  the neutrino physics \cite{Enqvist}.   The most stringent bound
$\lambda'_{131}
  \stackrel{<}{\sim} 0.3$  comes from the atomic parity violation experiment
  \cite{Barger}.

  As mentioned in 2.3.3, only two-body decay modes $(a),(b)$ and $(h)$ are
  targets of our study.   The formulae of the decay width for each mode are
  respectively given by

\begin{eqnarray}
 && {\Gamma}    (\widetilde{t}_1 \rightarrow  t \widetilde{Z}_i  ) =
\frac{\alpha}{2 m^{3}_{\widetilde{t}_1}} \lambda^{\frac{1}{2}}
 (m^{2}_{\widetilde{t}_1}, m^{2}_t, m^{2}_{\widetilde{Z}_i})  \nonumber\\
&&  {{\times}}   \Big[(|F_L|^2+|F_R|^2)(m^{2}_{\widetilde{t}_1}- m^{2}_t -
m^{2}_{\widetilde{Z}_i})- 4m_t  m_{\widetilde{Z}_i} Re(F_R  F^{*}_L )\Big]
,
\\
&&F_L  \equiv    \frac{m_t N'^{*}_{i4}  \cos\theta_t} {2m_W \sin\theta_W
\sin\beta}
                                  +e_u(N'^{*}_{i1} -\tan\theta_W
N'^{*}_{i4})  \sin\theta_t ,          \\
&&F_R\equiv   \Bigl(e_u  N'_{i1} + \frac{1/2 - e_u \sin^2 \theta_W}
{\cos\theta_W \sin\theta_W} N'_{i2}\Bigr)  \cos\theta_t -  \frac{m_t
N'_{i4} \sin
\theta_t}{2m_W  \sin\theta_W \sin\beta},
\end{eqnarray}

\begin{eqnarray}
&& {\Gamma}    (\widetilde{t}_1  \rightarrow  b\widetilde{W}_k )=
        \frac{\alpha }{4\sin^2 \theta_W  m{3 \atop {\widetilde{t}_1}}}
        \lambda^{\frac{1}{2}}
       \Bigl(m^{2}_{\widetilde{t}_1}, m^{2}_b, m^{2}_{\widetilde{W}_k} \Bigr)
       \nonumber\\
&&    {\times}     \Big[(|G_L|^2+|G_R|^2)(m^{2}_{\widetilde{t}_1}- m^{2}_b-
m^{2}_{\widetilde{W}_k} )- 4m_b m_{\widetilde{W}_k}Re(G_R G^{*}_L )\Big],
            \\
&&G_L   \equiv     - \frac{m_b U^{*}_{k2} \cos\theta_t} {\sqrt{2}m_W
\cos\beta},       \\
&&G_R  \equiv  V_{k1} \cos\theta_t+  \frac{m_t  V_{k2} \sin\theta_t}
{\sqrt{2}m_{W} \sin \beta},
\end{eqnarray}
and
\begin{eqnarray}
{\Gamma}   (\widetilde{t}_1 \rightarrow  ed)= \frac{\lambda'^{2}_{131}}{16\pi}
\cos^2\theta_t  m_{\widetilde{t}_1},
\end{eqnarray}
where $\lambda(x,y,z)=x^2+y^2+z^2-2xy-2yz-2zx$.  $N'_{ij}, V_{kl}$ and $U_{kl}$
respectively stand for the neutralino and chargino mixing angles
\cite{Nilles} .
The mixing angles as well as  masses  of the neutralinos
$m_{\widetilde{Z}_i}$
and the charginos  $m_{\widetilde{W}_k}$ are determined from the basic
parameters
in the MSSM($\mu ,  \tan \beta, M_2$).   As seen from Fig.~13 the branching
ratio of
the stop depends on the stop mass.   If the stop is heavy enough, i.e.,
$m_{\widetilde{t}_1}>m_t+m_{\widetilde{Z}_i}$ or
$m_{\widetilde{t}_1}>m_b+m_{\widetilde{W}_k}$ and the RB coupling is
comparable with the gauge or Yukawa coupling
$ \lambda' {2 \atop  {131}}/{4\pi}\stackrel{<}{\sim} \alpha,\alpha_t $
there is a parameter region where
$BR(\widetilde{t}_1 \rightarrow t\widetilde{Z}_i)$ or
$BR(\widetilde{t}_1 \rightarrow b\widetilde{W}_k)$
competes with $BR(\widetilde{t}_1 \rightarrow ed)$.
For light $\widetilde{t}_1$,  however, the decay
$\widetilde{t}_1 \rightarrow ed$ predominates over other decay
channels and we have $BR(\widetilde{t}_1 \rightarrow ed)\simeq 100\%$.

{\bf (A)} The case of light stop : $m_{\widetilde{t}_1}<m_t
+m_{\widetilde{Z}_i} $
or $m_{\widetilde{t}_1}
 <m_b +m_{\widetilde{W}_k} $

   Assuming the stop has  $BR(\widetilde{t}_1 \rightarrow ed)\simeq 100\%$,
   we can calculate the inclusive differential cross section for $e^\pm p
   \rightarrow   e^\pm qX$ with polarized $e^\pm$ beams(see Fig.~12) as
follows:

\begin{eqnarray}
\frac{d\sigma}{dxdQ^2}[e^{\pm}_{L,R}]=\frac{2\pi\alpha^2}{x^2s^2}
\sum_{q}[q(x,Q^2)\sum_{i=1}^{4}T_i(e^{\pm}_{L,R} q)+
       \bar{q}(x,Q^2)\sum_{i=1}^{4}T_i(e^{\pm}_{L,R} q)],
\label{rbcr}
\end{eqnarray}
where the coefficients $T_i(e{\pm}_{L,R} q)$ are represented in the Appendix.
  For  $e^{-}_L  \bar {d} \rightarrow \widetilde{t}_1$ followed by the decay
  $\widetilde{t}_1
 \rightarrow e ^{-}_L  \bar{d} $ we have

\begin{eqnarray}
\bar {T}^{d }_4 [e^{-}_L] = \frac{1}{4}F^{2}_{RB} (\widetilde{t}_1)
\frac{s^2 x^2}{(sx-m^{2}_{\widetilde{t}_1})+m^{2}_{\widetilde{t}_1}
{\Gamma}  ^{2}_{\widetilde{t}_1} },
\end{eqnarray}
where the decay width ${\Gamma}  (\widetilde{t}_1  \rightarrow ed)$ is
given by (45).
{}From Eq.(47) we can expect a clean signal of the stop as a sharp
resonance peak in the distribution of the Bjorken parameter $x$.
The  position peak corresponds  to $x=m^{2}_{\widetilde{t}_1}/s$ for fixed
$Q^2$.
Figure 14 clearly shows the resonance behavior of the production process.
The background curve represents the predicton of SM.
It is obviously seen from Fig.~14 that the lower $Q^2$ cuts would be very
efficient to suppress the background expected from SM, since the $s$-channel
resonance contribution is independent of $Q^2$.
It would be worthy of remarking that the similar peak could be expected
in the leptoquark production at HERA\cite{Schrempp}.
We should point out that the stop with the RB couplings will be
discriminated from most of the leptoquarks by its distinctive properties:(1)
the $x$ peak originated from the stop would exist only in the NC (not exist
in the CC)
process due to  no RB stop couplings to neutrinos,
(2) $e^+$ beams are more favorable than $e^-$ beams as will be mentioned
later.
One of the leptoquarks $\widetilde{S}_{1/2}$ with the charge $Q=-\frac{2}{3}$
will give the same signature as the RB stop, if the stop decays into the
electron
and $d$-quark with $BR(\widetilde{t}_1 \rightarrow ed)\simeq 100\%$.
\footnote{H1 collabaration at HERA has given the lower mass bound
$m_{\widetilde{t}_1}\stackrel{>}{\sim}98$ GeV on the RB stop from
the negative result for the leptoquark $\widetilde{S}_{1/2}$ search at 95\% CL
or $\lambda'_{131} = 0.3$ \cite{Abt}.} The event rate depends on not
only the RB coupling strength $\lambda'_{131}  $ but also the kind of beams.
Figure 15 shows the $y$ distribution at fixed $x$ for $e^-$  and $e^+$ beams.
We can see that the $e^+$ beam is more efficient than the $e^-$ one to
distinguish the stop signal from the SM background.
 This can be understood from the fact that the $e^+$ collides with
 valence $d$-quark in the proton, while the $e^-$ does only with
 sea $\bar{d}$-quarks.   The difference of their distribution functions
 is naturally reflected in the cross sections in Fig.~15.
 It is hopefully expected that the longitudinally polarized $e^+$ and $e^-$
 beams will soon be available at HERA.   They could be advantageous to
 suppress the SM background.   In Fig.~16 we show the $y(=Q^2/s)$ dependence
 of the asymmetries defined by

\begin{eqnarray}
         C_R      \equiv       \frac{d\sigma(e^{+}_R)/dxdy-
d\sigma(e^{-}_R)/dxdy}
                                                {d\sigma(e^{+}_R)/dxdy+
d\sigma(e^{-}_R)/dxdy}
\end{eqnarray}
and
\begin{eqnarray}
        A_{e^{-}}    \equiv       \frac{d\sigma(e^{-}_L)/dxdy -
d\sigma(e^{-}_R)/dxdy}
{d\sigma(e^{-}_L)/dxdy+ d\sigma(e^{-}_R)/dxdy}.
\end{eqnarray}
 It will be seen that the longitudinally polarized $e^\pm$ beams will be useful
 to identify the stop signal for a reasonable range of $y$ in $C_R$ .
 Finaly we show the searchable parameter region in
 ($\lambda'_{131}, m_{\widetilde{t}_1}$) plane at HERA.
 In Fig.~17 the shaded region is experimentally excluded from
 the atomic parity violation experiment\cite{Barger}.
 The area inside the solid contour is accessible at HERA
 whose production rate is more than ten signal
 events above the SM background with $Q^2>10^3$ GeV$^2$ in 100 pb$^{-1} $
running.
{}From Fig.~17 it is seen that the stop mass reach is
\begin{eqnarray}
    m_{\widetilde{t}_1}\stackrel{<}{\sim}200(270) {\rm GeV}          \nonumber
\end{eqnarray}
for $e^-(e^+)$ beams at the RB coupling
\begin{eqnarray}
    \lambda'_{131}  \simeq 0.1
    \nonumber
\end{eqnarray}

{\bf (B)} The case of heavy stop: $m_{\widetilde{t}_1}>m_t
+m_{\widetilde{Z}_i} $
or $m_{\widetilde{t}_1}>m_b +m_{\widetilde{W}_k} $

If the mass of the stop is heavy enough, various decay channels as $(a)$
to$ (g)$
compete with $(h)$ and a sharp peak at $x={m_{\widetilde{t}_1}}^2/2$ can no
longer be expected.   For the case of $BR(\widetilde{t}_1 \rightarrow
ed)\ll 100\%$,
 we should take into account of the processes
\begin{equation}
             ep         \rightarrow     t  \widetilde{Z}_i X
\label{tn}
\end{equation}
and
\begin{equation}
           ep         \rightarrow      b  \widetilde{W}_k X
\label{bc}
\end{equation}
as mentioned before.   Their Feynman diagrams are shown in Fig.~18.
Here we should consider the virtual contributions of the selectron,
sneutrino and d-squark with the same RB couplings constants$ \lambda'_{131} $.
 The differential cross sections are given by
\begin{eqnarray}
&&\frac{d\sigma}{dxdQ^2}(ep \rightarrow t\widetilde{Z}_i X)   =
\frac{\alpha \lambda '^{2}_{131}}{8 \hat{s}^2}
\Big[ |F_{\widetilde{e}}|^2 \frac{(\hat{u}-m^{2}_t)
(\hat{u}-m^{2}_{\widetilde{Z}_i})}     {(\hat{u}-m^{2}_{\widetilde{e}_L})^2} +
           |F_{\widetilde{d}}|^2 \frac{(\hat{t}-m^{2}_t)
(\hat{t}-m^{2}_{\widetilde{Z}_i})}
{(\hat{t}-m^{2}_{\widetilde{d}_R})^2}  \nonumber\\
&&  + \frac{\cos^2 \theta _t \hat{s}}
{(\hat{s}-m^{2}_{\widetilde{t}_1})^2-m^{2}_{\widetilde{t}_1}{\Gamma}
^{2}_{\widetilde{t}_1}}
\Bigl((|F_L|^2+|F_R|^2) (\hat{s}-m^{2}_t -m^{2}_{\widetilde{Z}_i})
-4m_t m_{\widetilde{Z}_i}Re(F_R F{*}_L)\Bigr)                \nonumber\\
&&  +2Re(F_{\widetilde{e}}F^{*}_{\widetilde{d}})
\frac{\hat{t}\hat{u}-m^{2}_t m^{2}_{\widetilde{Z}_i}}
{(\hat{u}-m^{2}_{\widetilde{e}_L})  (\hat{t}-m^{2}_{\widetilde{d}_R})}
   \nonumber\\
&&   + \frac{2\cos^2 \theta _t \hat{s}(\hat{s}-m^{2}_{\widetilde{t}_1})}
{((\hat{s}-m^{2}_{\widetilde{t}_1})^2 +m^{2}_{\widetilde{t}_1}
{\Gamma}  ^{2}_{\widetilde{t}_1} )
(\hat{u}-m^{2}_{\widetilde{e}_L})}
Re\Bigl(F^{*}_{\widetilde{e}}(F_R\hat{u}+F_Lm_tm_{\widetilde{Z}_i})\Bigr)
\nonumber\\
  && + \frac{2\cos^2 \theta _t \hat{s}(\hat{s}-m^{2}_{\widetilde{t}_1})}
{((\hat{s}-m^{2}_{\widetilde{t}_1})^2 +m^{2}_{\widetilde{t}_1}
{\Gamma}^{2}_{\widetilde{t}_1} )
(\hat{u}-m^{2}_{\widetilde{d}_R})}
Re\Bigl(F^{*}_{\widetilde{d}}(F_R\hat{t}+F_Lm_tm_{\widetilde{Z}_i})\Bigr)
\Big],
\end{eqnarray}
where $\hat{s}=xs, \hat{t}=-Q^2$ and
\begin{eqnarray}
&& F_{\widetilde{e}} \equiv e_e N'_{i1} - \frac{1/2+e_e\sin^2\theta
_W}{\cos\theta_W\sin\theta_W}N'_{i2},
               \nonumber\\
&&F_{\widetilde{d}} \equiv e_e N'_{i1} -  e_d\tan\theta _W N'_{i2} .
\nonumber
\end{eqnarray}
\begin{eqnarray}
&&  \frac{d\sigma}{dxdQ^2}(ep \rightarrow b\widetilde{W}_k X)  =
\frac{\alpha \lambda '^{2}_{131}}{16 \hat{s}^2\sin^2\theta_W}
\Big[|V_{11}|^2  \frac{(\hat{u}-m^{2}_b)(\hat{u}-m^{2}_{\widetilde{W}_k})}
      {(\hat{u}-m^{2}_{\widetilde{\nu}})^2}                     \nonumber\\
 &&      + \frac{\cos^2 \theta _t \hat{s}}
{(\hat{s}-m^{2}_{\widetilde{t}_1})^2-m^{2}_{\widetilde{t}_1}
{\Gamma}  ^{2}_{\widetilde{t}_1}}\Bigl((|G_L|^2+|G_R|^2)
(\hat{s}-m^{2}_b -m^{2}_{\widetilde{W}_k})
-4m_b m_{\widetilde{W}_k}Re(G_R G{*}_L)\Bigr)          \nonumber\\
&&   - \frac{2\cos^2 \theta _t \hat{s}(\hat{s}-m^{2}_{\widetilde{t}_1})}
{\Bigl((\hat{s}-m^{2}_{\widetilde{t}_1})^2 +m^{2}_{\widetilde{t}_1}
{\Gamma}  ^{2}_{\widetilde{t}_1} \Bigr)(\hat{u}-{m^{2}_{\widetilde{\nu}}}  )}
Re\Bigl(V^{*}_{11}(G_R\hat{u}+G_Lm_b {m_{\widetilde{W}_k}}     )\Bigr)  \Big]
\end{eqnarray}

Also from Fig.~19 we can infer that $e^+$ beams are more efficient than
$e^-$ ones.
It is expected that the detectable cross sections $\sigma
\stackrel{>}{\sim}0.1$ pb
for heavy stop with mass ${m_{\widetilde{t}_1 }}\stackrel{<}{\sim} 250$ GeV for
$e^+$ beams.   As far as $e^-$ beams are concerned only
$e^-p \rightarrow b\widetilde{W}_k X$ would be detectable for
${m_{\widetilde{t}_1}} \stackrel{<}{\sim} 170$ GeV.
In our model the LSP, the lightest neutralino $\widetilde{Z}_1$
will decay into $R$-even particles via only non-zero RB coupling
$ \lambda'_{131} $.     A typical decay chain will be (see Fig.~20)
\begin{eqnarray}
       ep \rightarrow t\widetilde{Z}_1 X \rightarrow  (bW)(bd\nu) X
       \rightarrow (b(\ell\nu))(bd\nu)X  .
\end{eqnarray}
The chargino $\widetilde{W}_1$ will also decay into $R$-even particles through
$\widetilde{Z}_1$ like
\begin{eqnarray}
       ep \rightarrow b\widetilde{W}_1 X  \rightarrow
(b\ell\nu\widetilde{Z}_1) X
       \rightarrow b(\ell\nu(bd\nu))X.
\label{signature}
\end{eqnarray}
  In both processes (\ref{tn}) and (\ref{bc}) 2b-jets+jet+lepton+missing
  transverse momentum
 ${\ooalign{\hfil/\hfil\crcr$P$}}_T$ would emerge as final products to be
 detectable.
 Figure 21 shows the Monte Carlo events for the transverse momentum
 distribution of scattered muon from the process (\ref{signature}) under
 the condition of the integrated luminosity ${\cal L}=$ 300 pb$^{-1}$ for
$e^-p$
 collisions at HERA.
 The branching ratio $BR(\widetilde{W}_1\rightarrow \nu\mu\widetilde{Z}_1)$ is
 assumed to be $\frac{1}{9}$\cite{Schimert}.   Shown in Fig.~21 are
distributions for
 ${m_{\widetilde{t}_1}}=150$ GeV and ${m_{\widetilde{t}_1}}=$100 GeV together
 with background(short-dashed line) of muon events coming from charged current
 process(CC)  $ e^-p\rightarrow \nu qX$ and $W$-gluon fusion process (WGF)
  $ e^-p\rightarrow  \nu s\bar{c}X,  \nu b \bar{c}X$.  The generators
  LEPTO \cite{Ingelman} and AROMA\cite{IngelmanSch} with JETSET
\cite{Sjostrand}
  have , respectively, been used for CC and WGF.   From Fig.~19 we can see
that
  $ep \rightarrow  b\widetilde{W}_1  X$ will have a reasonably feasible cross
  section to which the stop contributes from the s-channel for
  ${m_{\widetilde{t}_1}}\stackrel{>}{\sim}$100 GeV.
  This is also the case for $ep \rightarrow  t\widetilde{Z}_1  X$
  since the same final states are realized.
  Thus, in both processes,
  $ep \rightarrow  t\widetilde{Z}_1  X$and $ep \rightarrow
b\widetilde{W}_1  X$
  a possible typical signature of the stop production would be
  2b-jets+jet+lepton+missing  transverse momentum
  ${\ooalign{\hfil/\hfil\crcr$P$}}_T$  owing to the LSP decay via RB
coupling.
  One of the signals to be detected at HERA is characterized by the high $P_T$
  spectrum of muons where the lower $P_T$ cut makes the event distinctive from
  its background.
  Although it is difficult to discriminate the stop from one of
  the leptoquark $\widetilde{S}_{1/2}$ for light enough stop
  $m_{\widetilde{t}_1}<m_t +m_{\widetilde{Z}_i},
  m_{\widetilde{t}_1}<m_b +m_{\widetilde{W}_k} $,
  a characteristic signature of the heavy stop with mass
   $m_{\widetilde{t}_1}>m_t +m_{\widetilde{Z}_i},
   m_{\widetilde{t}_1}>m_b +m_{\widetilde{W}_k} $
   could clearly  be distinguished from the leptoquark $\widetilde{S}_{1/2}$.

  Recent observation of a single muon of high $P_T$ event in
  $e^+ p \to \mu^+ X$ by H1 \cite{sgmu} gives us a favourable evidence for our
  arguments on the stop.
  As seen from Fig.19 a few events would be expected for
  $m_{\widetilde{t}_1}$ $=$ 150 GeV and present luminosity
  $L$ $\simeq$ 3 pb$^{-1}$.

\section{\it Concluding remarks}

We have discussed SUSY particle searches at HERA by paying our particular
attention upon the single production of stop in the framework of the MSSM
with $R$-parity breaking interactions.
Observation of the moderately light stop predicted by the model would be a
nice target of HERA experiments so far the RB coupling constant
$\lambda '_{131}$ $\nle$ 0.3.

In the case of BR($\widetilde{t}_1 \to ed$) $\simeq$ 100\% corresponding
to $m_{\widetilde{t}_1} < m_t +m_{\widetilde{Z}_1} $
or $m_{\widetilde{t}_1}
 < m_b +m_{\widetilde{W}_1} $,
the stop produced via neutral currents can clearly be seen as a resonance peak
at the Bjorken parameter $x$ distribution.
Heavier stop, $m_{\widetilde{t}_1} > m_t +m_{\widetilde{Z}_1} $
or $m_{\widetilde{t}_1}
 > m_b +m_{\widetilde{W}_1} $,
are produced through charged currents emerge from the following chain of
decay processes :
\begin{eqnarray}
       ep \rightarrow t\widetilde{Z}_1 X \rightarrow  (bW)(bd\nu) X
       \rightarrow (b(\ell\nu))(bd\nu)X  .
\end{eqnarray}
The chargino $\widetilde{W}_1$ will also decay into $R$-even particles through
$\widetilde{Z}_1$ like
\begin{eqnarray}
       ep \rightarrow b\widetilde{W}_1 X  \rightarrow
(b\ell\nu\widetilde{Z}_1) X
       \rightarrow b(\ell\nu(bd\nu))X.
\end{eqnarray}
The evidence for the existence of the stop would emerge from the careful
identification of 2$b$-jets $+$ jet $+$ $\ell$ $+$
${\ooalign{\hfil/\hfil\crcr$P$}}_T$.
Possible backgrounds come from
$e^- p \to \nu q X$ and $e^- p \to \nu s \bar{c} X$, $\nu b \bar{c} X$.
The transverse momentum distribution of
scattering muon will distinguish the stop events more clearly from
backgrounds.
Extensive Monte Carlo events with a variety of SUSY parameter sets would
highly be desirable.
Since the production cross section of the stop is unfortunately rather low
the high luminosity is very much favourable.

Until LEP{\uppercase\expandafter{\romannumeral 2}} or Next Liner Collider
will be built we are sure that HERA would play a unique role to open the
"SUSY world" through our stop with mass of 100 to 300 GeV.
TEVATRON groups are also enthusiastic about sparticle search at high
energies.
We are very much expecting HERA and TEVATRON complementarily step
forward to enter into the novel region of particle physics by making full
use of their specialities of detectors.


\vfill\eject

\begin{flushleft}
{\Large{\bf Appendix}}
\end{flushleft}

  The analytic expression for the cross section Eq.(\ref{rbcr})
is given as follows.
\begin{eqnarray}
{{\ddf\sigma}\over{\ddf x\ddf Q^{2}}}[e^{-}_{L,R}]
&=&{{2\pi\alpha^{2}}\over{x^{2}s^{2}}}
   \left[
\sum_{\i =1}^{2}\tui[e^{-}_{L,R}]\xq u\xq +
\sum_{\i =1}^{2}\tuib[e^{-}_{L,R}]\xq \bar{u}\xq \right. \\
&  & \left. +
\sum_{\i =1}^{4}\tdi[e^{-}_{L,R}]\xq d\xq +
\sum_{\i =1}^{4}\tdib[e^{-}_{L,R}]\xq \bar{d}\xq
  \right],
\end{eqnarray}
where
\begin{eqnarray}
\tuo[e^{-}_{L}]&=&{{(Q^2 -sx)^2}\over{Q^4}}
(\ee\eu +\qz\ael\aur)^2,\\
\tuo[e^{-}_{R}]&=&{{(Q^2 -sx)^2}\over{Q^4}}
(\ee\eu +\qz\aer\aul)^2,\\
\tut[e^{-}_{L}]&=&{{s^2 x^2}\over{Q^4}}
(\ee\eu +\qz\ael\aul)^2,\\
\tut[e^{-}_{R}]&=&{{s^2 x^2}\over{Q^4}}
(\ee\eu +\qz\aer\aur)^2,\\
\tuob[e^{-}_{L}]&=&{{s^2 x^2}\over{Q^4}}
(\ee\eu +\qz\ael\aur)^2,\\
\tuob[e^{-}_{R}]&=&{{s^2 x^2}\over{Q^4}}
(\ee\eu +\qz\aer\aul)^2,\\
\tutb[e^{-}_{L}]&=&{{(Q^2 -sx)^2}\over{Q^4}}
(\ee\eu +\qz\ael\aul)^2 ,\\
\tutb[e^{-}_{R}]&=&{{(Q^2 -sx)^2}\over{Q^4}}
(\ee\eu +\qz\aer\aur)^2 ,\\
\tdo[e^{-}_{L,R}]&=&\left.\tuo[e^{-}_{L,R}]\right|_{u \rightarrow d}, \qquad
\tdt[e^{-}_{L,R}]=\left.\tut[e^{-}_{L,R}]\right|_{u \rightarrow d},\\
\tdh[e^{-}_{L}]&=&-\frb(\stl)
    {{(Q^2 -sx)^{2}}\over{Q^2\utg}}
    (\ee\ed +\qz\ael\adr),\\
\tdh[e^{-}_{R}]&=&0,\\
\tdf[e^{-}_{L}]&=&{{1}\over{4}}\frb^{2}(\stl)
    {{(Q^2-sx)^2}\over{\utg^2}},\\
\tdf[e^{-}_{R}]&=&0,\\
\tdob[e^{-}_{L,R}]&=&\left.\tuob[e^{-}_{L,R}]\right|_{u \rightarrow d}, \qquad
\tdtb[e^{-}_{L,R}]=\left.\tutb[e^{-}_{L,R}]\right|_{u \rightarrow d},\\
\tdhb[e^{-}_{L}]&=&-\frb(\stl)
    {{s^2 x^2 (sx-\mstl^2)}\over{Q^2(\stg)}}
    (\ee\ed +\qz\ael\adr),\\
\tdhb[e^{-}_{R}]&=&0,\\
\tdfb[e^{-}_{L}]&=&{{1}\over{4}}\frb^{2}(\stl)
    {{s^2 x^2}\over{\stg}}, \\
\tdfb[e^{-}_{R}]&=&0
\end{eqnarray}
for $e^{-}$ beams.
The formula for $e^{+}$ beams can be obtained by
the following replacement in the above formula for the $e^{-}$ beams ;
\begin{eqnarray}
e^{-}_{L,R}&\rightarrow& e^{+}_{R,L} \\
q&\rightarrow& \bar{q} \\
\bar{q}&\rightarrow& q
\end {eqnarray}
Here, ${e}^{f}$ denote the electromagnetic charge of
matter fermion $f$, and
\begin{eqnarray}
A_{L}^{f} & \equiv &
-{{{T}_{3}^{f}-{e}^{f}\sin^{2}\theta_{W}}
                      \over{\cos\theta_{W}\sin\theta_{W}}}, \\
A_{R}^{f} & \equiv &
{e}^{f}\tan\theta_{W},
\end{eqnarray}
where ${T}_{3}^{f}$ are the third component of isospin and $\theta_{W}$
is the Weinberg angle.

\vfill\eject

\vfill\eject

{\Large{\bf Figure Captions}}
\begin{description}

\item[{\bf Figure 1:}]
 D0, LEP and UA1/UA2  squark and gluino
 mass limits as a function of squark and gluino mass\cite{Hagopian}.
\label{fig1}

\item[{\bf Figure 2:}]
The excluded regions of the MSSM parameter space at 95\% CL as a function of
the parameters $M$ and $\mu$.   The kinematical limit corresponds to the sum
of the lightest and next to lightest neutralino masses being equal to the
center-of-mass energy.   The exclusion coming from the lineshape measurement
goes beyond the kinematic limit of the direct search \cite{Acciarri}.
\label{fig2}

\item[{\bf Figure 3:}]
Neutralino lower mass limits(95\% CL). The lines correspond to the lower mass
limit for the different neutralinos, where ${\chi}$,  ${\chi}' $,
${\chi}''  $ and ${\chi}''' $ in this figure represent
$\widetilde {Z}_1$,  $\widetilde {Z}_2$,       $\widetilde {Z}_3$ and
$\widetilde {Z}_4$  in the text, respectively.  The regions below the lines
are excluded \cite{Acciarri}.
\label{fig3}

\item[{\bf Figure 4:}]
 Excluded region   on   the ($m_{\widetilde {\gamma}}$, $m_{\widetilde {e}}$)
 plane.
VENUS (90\% CL, solid line), ASP (90\% CL, dashed line),
CELLO (90\% CL, dotted line),  ALEPH (95\% CL, dot-dashed line)
and combined single photon result (90\% CL, solid line) \cite{Hosoda}.
\label{fig4}

\item[{\bf Figure 5:}]
 The region in the ($\mu, M_2$ ) plane  for $\tan \beta =2$ excluded by the
 experimental data  on
\begin{description}
\item[A] lower bound on the mass of lighter chargino $m_{\widetilde {W}_1} >
45$ GeV,
\item[B]upper bound on the branching ratio of the visible neutralino mode
$Br(Z^0\rightarrow visi.) < 5\times10^{-6}$,
\item[C] upper bound on the invisible width of the $Z^0$  , and
\item[D] accepted gluino mass bound at CDF $m_{\widetilde {g}} >150$ GeV
(90\% CL),
\end{description}
where the hatched regions  of each contour have been excluded\cite{Kon}.
\label{fig5}

\item[{\bf Figure 6:}]
 The excluded region  in the ($\theta_t (\vartheta_{mix}), m_{\widetilde
{t}_1} $)
 plane  at 95\% CL,  for the case of the mass difference  $\Delta m$
 between  ${\widetilde {t}_1}$ and ${\widetilde {Z}_1}$  larger than 5 GeV.
 The region excluded from the limit on the $Z^0$ decay width
 ($\Delta\Gamma\le 26$ GeV at 95\% CL) and limits
 from previous publications are also shown \cite{Akers}.
\label{fig6}

\item[{\bf Figure 7:}]
 The excluded regions  in the ($m_{\widetilde {t}_1}, m_{\widetilde {Z}_1}$)
 plane at 95\% CL, where the mixing angle is assumed to be
 $\theta_t (\vartheta_{mix}) \le 0.85 $ or $\ge 1.15$ rad (shaded area),
 and  $\theta_t (\vartheta_{mix}) \le 0.97 $ or $\ge 0.99$ rad (hatched area).
 The dashed curve shows the contour of the limit from the previonus
 publication \cite{Akers}.
\label{fig7}

\item[{\bf Figure 8:}]
 Rejection  limits at the 95\% CL for the coupling $\lambda '_{111}$
 as a function of the squark mass for various fixed photino masses.
 Regions above the curves are excluded. The limits combine all charged
 and neutral decays of the $\widetilde {d}$ and
$\widetilde {\bar{u}}$ \cite{Kohler}.
\label{fig8}

\item[{\bf Figure 9:}]
 The 95\% CL upper limits on the couplings of squarks versus mass for (a)
 the  $\widetilde {d}$
(in order of decreasing coupling limits based on the NC, CC and combined
samples) and (b) the $\widetilde {\bar{u}}$ and
$\widetilde {\bar{t}}$ using the NC samples only.
The sensitivity of the result to the photino mass is also shown \cite{McLean}.
\label{fig9}

\item[{\bf Figure 10:}]
 Feynman diagrams for $ e^- g  \rightarrow  e^-
  \widetilde{t}_1   \widetilde{t}^{*}_1$ \cite{stopbg}.
\label{fig10}

\item[{\bf Figure 11:}]
Total cross-sections  for  $ e^- p  \rightarrow  e^-   \widetilde{t}_1
\widetilde{t}^{*}_1  X$  with  cut $Q^2$ $>$ 5 GeV$^2$.
Solid line, dashed line and dotted line correspond to exact calculation
including both $\gamma$ and $Z$, including only $\gamma$ and
WWA, respectively \cite{stopbg}.
\label{fig11}

\item[{\bf Figure 12:}]
 Feynman diagrams for sub-processes $e^\pm q \rightarrow e^\pm q$
\cite{sthera}.
\label{fig12}

\item[{\bf Figure 13:}]
 $ m_{\widetilde{t}_1}$ dependence of branching ratio of stop. We take
 $m_t $= 135 GeV, $\tan\beta$=2, $\theta_t$=1.0, $\lambda '_{131}$=0.1 and
 ($M_2$ (GeV), $\mu$ (GeV)) = (50, $-$100) for (a) and (100, $-$50) for (b)
 \cite{stopsg}.
\label{fig13}

\item[{\bf Figure 14:}]
 $x$ distribution at fixed $Q^2$. Adopted parameters are
 $ m_{\widetilde{t}_1}$=200 GeV, $\lambda '_{131}$=0.25 and $\theta_t$=0.0
 \cite{sthera}.
\label{fig14}

\item[{\bf Figure 15:}]
 $y$ distributions at fixed $x$ using the electron  and the positron.
 $x$ is fixed at 0.2.  Adopted parameters are $ m_{\widetilde{t}_1}$=140 GeV
 and $\theta_t$=$\pi/4$ \cite{sthera}.
\label{fig15}

\item[{\bf Figure 16:}]
 $y$ distribution at fixed $x$ of differential asymmetries $C_R $ (a) and
 $A_{e^-}$ (b). $x$ is fixed at 0.2. Adopted parameters are
 $ m_{\widetilde{t}_1}$=140 GeV and $\theta_t$=0 \cite{sthera}.
\label{fig16}

\item[{\bf Figure 17:}]
 Searchable parameter region at HERA in  ($\lambda '_{131}, m_t $).
 Kinematical cut is  $Q^2$ $>$ $10^3$ GeV$^2$ \cite{sthera}.
\label{fig17}

\item[{\bf Figure 18:}]
  Feynman diagrams for  sub-processes
  $ ep    \rightarrow     t \widetilde{Z}_i$ and
 $ ep  \rightarrow      b  \widetilde{W}_k$ \cite{stopsg}.
\label{fig18}

\item[{\bf Figure 19:}]
Stop mass dependence of total cross section.
We take $m_t$ =135 GeV, $\theta_t$ = 1.0, tan$\beta$ = 2,
$\lambda '_{131}$ = 0.1, $M_2$ = 100 GeV,  $m_{\tilde{\ell}}$ = 200 GeV,
             $m_{\tilde{q}}$ = 300 GeV and $\mu$ = $-50$ GeV.   Solid,
             short-dashed, dotted and dashed lines correspond to
             $ e^- p  \rightarrow      b  {\widetilde{W}_1}^-  X$,
             $e^+ p  \rightarrow  \bar{b}  {\widetilde{W}_1}^+  X$,
             $e^- p  \rightarrow \bar{t} \widetilde{Z}_1 X$ and
             $e^+ p   \rightarrow  t \widetilde{Z}_1 X$ , respectively
\cite{stopsg}.
\label{fig19}

\item[{\bf Figure 20:}]
 Feynman diagrams for  LSP decay \cite{stopsg}.
\label{fig20}

\item[{\bf Figure 21:}]
 Monte Carlo events for transverse momentum distribution of scattered muon
from    $ e^- p  \rightarrow    b  \widetilde{W}_1  X$ (solid lines)
 together with backgrounds CC and WGF processes (short-dashed line).
 We take  $m_{\tilde{\ell}}$=200 GeV,   $m_{\tilde{q}}$=300 GeV ,
 $m_t$=135 GeV,   $\theta_t$=1.0, tan$\beta $=2,
  $\lambda '_{131}$=0.1, $M_2$=100 GeV,
  $\mu$=$-$50 GeV and integrated luminosity $L$=300 pb$^{-1}$ \cite{stopsg}.
\label{fig21}

\end{description}

\vfill\eject

\begin{center}
{{\bf Table {\uppercase\expandafter{\romannumeral 1}}}} \quad
{Particle content}\\
\end{center}
\begin{center}
\begin{tabular}{|c|c|c|c|c|c|c|c|c|}
\hline
Superfields & Spin-0 & Spin-$1/2$ & Spin-1 & Color & $T$ & $Y$ & $B$ & $L$ \\
\hline
$\hat{Q}$ & $\sql$            & $q_L$       &   & 3     & $\half$ &
$+{\frac{1}{6}}$
& $+{\frac{1}{3}}$ & 0 \\
$\hat{U^c}$ & $\overline{\sur}$ & ${(u^c)}_L$ &   & 3$^*$ & 0       &
$-{\frac{2}{3}}$
& $-{\frac{1}{3}}$ & 0 \\
$\hat{D^c}$ & $\overline{\sdr}$ & ${(d^c)}_L$ &   & 3$^*$ & 0       &
$+{\frac{1}{3}}$
& $-{\frac{1}{3}}$ & 0 \\
$\hat{L}$ & $\sll$            & $\ell_L$    &   & 1     & $\half$ &
$-{\frac{1}{2}}$ & 0
& $+1$ \\
$\hat{E^c}$ & $\overline{\ser}$ & ${(e^c)}_L$ &   & 1     & 0       & $+1$
           & 0
& $-1$ \\
\hline
$\hat{G}$ &        & $\gluino$ & $g$ & 8 & 0 & 0 & 0 & 0 \\
$\hat{W}$ &       & $\wino$   & $W$ & 1 & 1 & 0 & 0 & 0 \\
$\hat{B}$ &        & $\bino$   & $B$ & 1 & 0 & 0 & 0 & 0 \\
\hline
$\hat{H_1}$ & $H_1$ & $\widetilde{H_{1L}}$ &   & 1 & $\half$ & $-{\frac{1}{2}}$
& 0 & 0 \\
$\hat{H_2}$ & $H_2$ & $\widetilde{H_{2L}}$ &   & 1 & $\half$ &
$+{\frac{1}{2}}$ &
0 & 0 \\
\hline
\end{tabular}
\label{table1}
\end{center}


\begin{thebibliography}{99}

\bibitem{dp}
S. Dimopoulos, {\it Proc. of the {\uppercase\expandafter{\romannumeral 27}}th
Int. Conf. on High Energy Physics, Glasgow, 1994},
ed. P. J. Bussey and I. G. Knowles,
 (Institute of Physics Publishing, Bristol and Philadelphia, 1995), p.93
\bibitem{top}
F. Abe et al. (CDF Collab.), {\it Phys. Rev. Lett.} {\bf 74}, 2626 (1995) ;
S. Abachi et al. (D0 Collab.), {\it Phys. Rev. Lett.} {\bf 74}, 2632 (1995)
\bibitem{Hagiwara}
K. Hagiwara, talk at {\it Int. Symposium on Lepton and Photon Interactions
at High Energies}, Beijing, 1995
\bibitem{unif}
H. Georgi and S. Glashow, {\it Phys. Rev. Lett.} {\bf 32}, 438 (1974) ;
J. C. Pati and A. Salam,  {\it Phys. Rev. } {\bf D8}, 1240 (1973)
\bibitem{tech}
S. Weinberg,  {\it Phys. Rev.} {\bf D13}, 974 (1979) ; {\bf D19}, 1277 (1979) ;
L. Susskind, {\it Phys. Rev.} {\bf D20}, 2619 (1979) ;
E. Fahri and L. Susskind, {\it Phys. Rep.} {\bf 74}, 277 (1981)
\bibitem{Nilles}
  For reviews, see,
H. Nilles, {\it Phys. Rep. } {\bf 110}, 1 (1984) ;
H. Haber and G. Kane, {\it Phys. Rep. } {\bf 117},  75  (1985)
\bibitem{sst}
M. B. Green, J. H. Schwartz and E. Witten,{\it  "Superstring Theory
{\uppercase\expandafter{\romannumeral 1}},
{\uppercase\expandafter{\romannumeral 2}}" },
(Cambridge University Press, 1987)
\bibitem{running}
  U. Amaldi et al., {\it Phys. Rev.} {\bf D36}, 1385 (1987);
  J. Ellis S. Kelley and D.V. Nanopoulos, {\it Phys. Lett.} {\bf B249},
 441 (1990);
  P. Langacker and M. Luo, {\it Phys. Rev.} {\bf D44},
 817 (1991) ;
  F. Anselmo et al., {\it Nuovo Cimento} {\bf 104A},
(1991) 1817; {\bf 105A} 1357 (1992);
  A. Zichichi, Talk at {\it 10 Yeras of SUSY Confronting Experiment},
  Geneva,Switzerland, 7-9 ,Sep.,1992
\bibitem{dm}
For example, see, L. Roszkowski, "{\it Supersymmetric Dark Matter - A Review}",
{\it Proceedings of 23rd Workshop of the INFN Eloisatron Project, Erice, 1992},
  ed. L.Cifarelli and V.A.Khoze,  (World Scientific, Singapore,1993), p. 429
\bibitem{stop}
 J. Ellis and S. Rudaz, {\it Phys. Lett.} {\bf 128B}, 248  (1983);
G. Altarelli and R. R\"uckl, {\it Phys. Lett.} {\bf 144B}, 126  (1984);
I. Bigi and S. Rudaz, {\it Phys. Lett.} {\bf 153B}, 335  (1985)
\bibitem{HK}
  K. Hikasa and M. Kobayashi, {\it Phys. Rev.} {\bf D36}, 724 (1987)
\bibitem{Rb}
J. D. Wells, C. Kolda and G. L. Kane, {\it Phys. Lett.} {\bf B338}, 219
(1994) ;
D. Garcia, R. Jimenez and J. Sola, {\it Phys. Lett.} {\bf B347}, 321 (1995) ;
X. Wang, J. L. Lopez and D. V. Nanopoulos, CTP-TAMU-25/95 (hep-ph/9506217)
\bibitem{stoprb}
  T. Kon and T. Kobayashi, {\it Phys. Lett.} {\bf B270}, 81 (1991)
\bibitem{susana}
 J. Butterworth and H. Dreiner,
{\it Proc. of the HERA Workshop : "Physics at HERA"} 1991,
eds. W. Buchm\"uller and G. Ingelman,
Vol.2, p.1079 ; {\it Nucl. Phys.} {\bf B397}, 3 (1993)
\bibitem{Hewett}
J. L. Hewett,
{\it "Research Directions for the Decade",
Proc. of 1990 Summer Study on High Energy Physics, Snowmass, 1990},
ed. E. L. Berger, (World Scientific, Singapore, 1992), p.566
\bibitem{Barger}
V. Barger, G. F. Giudice and T. Han, {\it Phys. Rev.} {\bf D40}, 2987 (1989)
\bibitem{bartl}
A. Bartl et al.,
{\it Proceedings of the HERA Workshop : Physics at HERA},
W. Buchm\"{u}ller and G. Ingelman (eds), Vol2, p.1118 (1991)



\bibitem{higgs}
J. Gunion and H. Haber, {\it Nucl. Phys.} {\bf B272}, 1 (1986)
\bibitem{susy}
  For example, H. E. Haber, {\it Note on Supersymmetry} in
Review of Particle properties, Part
{\uppercase\expandafter{\romannumeral 2}},
{\it Phys. Rev.} {\bf D45},
{\uppercase\expandafter{\romannumeral 9}}.5  (1992)
\bibitem{RGE}
  I. Inoue, A. Kakuto, H. Komatsu and S. Takeshita,
{\it Prog. Theor. Phys.} {\bf 67} (1982) 1889 ; {\bf 68}, 927  (1982);
J. Ellis, J. Hagelin, D. Nanopoulos and K. Tamvakis,
{\it Phys. Lett.} {\bf 125B}, 275  (1983);
L. Alvarez-Gaum\'e, J. Polchinski and M. Wise, {\it Nucl. Phys.}
{\bf B221}, 495 (1983) ;
L. Ib\'a\~nez  and C. L\'opez, {\it Phys. Lett.}
{\bf 126B}, 54  (1983)
\bibitem{Hikasa}
  K. Hikasa, {\it "JLC Supersymmetry Manual"}, unpublished
\bibitem{DH}
  M. Drees and K. Hikasa, {\it Phys. Lett.} {\bf B252}, 127 (1990)




\bibitem{Dreiner}
  H. Dreiner, {\it Proceedings of 23rd Workshop of the INFN Eloisatron
Project},
  ed. L.Cifarelli and V.A.Khoze,  (World Scientific, Singapore,1993), p.195
\bibitem{Fayet}
 P. Fayet, {\it Proceedings of 23rd Workshop of the INFN Eloisatron Project},
 ed. L.Cifarelli and V.A.Khoze,  (World Scientific, Singapore,1993), p.1
\bibitem{Komamiya}
S. Komamiya. Invited talk at the Recontres du Vietnam, Hanoi  Vietnam, 1993
(Preprint  UT-ICEPP 94-04, 1994)
\bibitem{Hu}
 Ping Hu, {\it Proceedings SUSY93 Workshop},  ed. P. Nath,
 (World Scientific, Singapore,1994), p.229
\bibitem{Hagopian}
 S. Hagopian, {\it Proc. of the {\uppercase\expandafter{\romannumeral 27}}th
Int. Conf. on High Energy Physics, Glasgow, 1994},
 ed. P. J. Bussey and I. G. Knowles,
 (Institute of Physics Publishing, Bristol and Philadelphia, 1995), p.809
\bibitem{Zhou}
 B. Zhou , {\it Proceedings SUSY93 Workshop},  ed. P. Nath,
 (World Scientific, Singapore,1994), p.54
\bibitem{Medcalf}
 T. Medcalf , {\it Proceedings SUSY93 Workshop},
 ed. P. Nath,  (World Scientific, Singapore,1994), p.33
\bibitem{Lutz}
 P. Lutz, {\it Proceedings SUSY93 Workshop},  ed. P. Nath,
 (World Scientific, Singapore,1994), p.45
\bibitem{Acciarri}
L3 Collab. M. Acciarri et al., CERN-PPE/95-14 (1995)
\bibitem{Abe}
TOPAZ Collab.,  T. Abe,  {\it Proceedings of the 2nd Workshop on TRISTAN
Physics
at High Luminosity}, ed. H.Sagawa, T. Tsukamoto, I. Watanabe and Y. Yamada,
(KEK, Tsukuba,   1994), p.418
\bibitem{Hosoda}
VENUS Collab.,  N. Hosoda  et al., {\it Phys. Lett.} {\bf B311}, 211 (1994)
\bibitem{Kon}
 T. Kon  and T. Nonaka, Seikei University preprint, ITP-SU-94/02 (1994)
\bibitem{Akers}
OPAL Collab., R.Akers et al., {\it Phys. Lett.} {\bf B337},  207 (1994)
\bibitem{Hultqvist}
K. Hultqvist, {\it Proceedings of the International Europhysics Conference on
High Energy Physics, Marseille, 1993}, ed. J. Carr and M. Perrottet,
(Editions Frontieres, Gif-sur-Yvette Cedex-France, 1994), p.276
\bibitem{Shirai}
VENUS Collab., J. Shirai ,  {\it Proceedings of the 2nd Workshop on TRISTAN
Physics
at High Luminosity}, ed. H.Sagawa, T. Tsukamoto, I. Watanabe and Y. Yamada,
(KEK, Tsukuba,   1994), p.405
\bibitem{EnomotoPR}
TOPAZ Collab., R. Enomoto et al., {\it Phys. Rev.}  {\bf D50}, 1879 (1994)
\bibitem{Enomoto}
TOPAZ Collab.,  R. Enomoto et al., {\it Proceedings of the 2nd Workshop on
TRISTAN Physics at High Luminosity}, ed. H.Sagawa, T. Tsukamoto, I.
Watanabe and
Y. Yamada, (KEK, Tsukuba,   1994), p.369
\bibitem{Okada}
 Y. Okada,  {\it Phys. Lett.} {\bf B315}, 119 (1993)
\bibitem{Fukugita}
 M. Fukugita, H. Murayama, M. Yamaguchi and T. Yanagida , {\it Phys. Rev.
Lett.}
 {\bf 72}, 3009 (1994)
\bibitem{Baer}
H.Baer, J. Sender and X.Tata, {\it Phys. Rev.} {\bf D50}, 4517 (1994)
\bibitem{Butterworth}
 J. Butterworth and H. Dreiner, {\it Nucl. Phys.} {\bf B397}, 3 (1993)
\bibitem{Enqvist}
K. Enqvist, A. Masiero and A. Riotto, {\it Nucl. Phys.} {\bf   B373}, 95(1992)
\bibitem{Abt}
 H1 Collab.,  I. Abt et al., {\it Nucl. Phys.} {\bf  B396}, 3 (1993)
\bibitem{Kohler}
 T. K\"{o}hler, {\it  Proc. of the {\uppercase\expandafter{\romannumeral 27}}th
Int. Conf. on High Energy Physics},
 ed. P. J. Bussey and I. G. Knowles, (Institute of Physics Publishing,
 Bristol and Philadelphia, 1995), p.801
\bibitem{McLean}
  ZEUS Collab., K. W. McLean , {\it Proceedings of the International
Europhysics
  Conference on High Energy Physics, Marseille, 1993}, ed. J. Carr and M.
Perrottet,
  (Editions Frontieres, Gif-sur-Yvette Cedex-France, 1994), p.268


\bibitem{susync}
S. K. Jones and C. H. Llewellyn Smith, {\it Nucl. Phys.} {\bf B217}, 145
(1983) ;
P. R. Harrison, {\it Nucl. Phys.} {\bf B249}, 704 (1985) ;
J. A. Bagger and M. E. Peskin, {\it Phys. Rev.} {\bf D31}, 2211 (1985) ;
J. Bartels and W. Hollik, {\it Z. Phys.} {\bf C39}, 433 (1988) ;
H. Komatsu and R. R\"uckl, {\it Nucl. Phys.} {\bf B299}, 401 (1988) ;
A. Bartl, H. Fraas and W. Majerotto, {\it Nucl. Phys.} {\bf B297}, 479
(1988) ;
{\it Z. Phys.} {\bf C41}, 475 (1988)
\bibitem{ncour}
T. Kon, K. Nakamura and T. Kobayashi, {\it Z. Phys.} {\bf C45}, 567 (1990)
\bibitem{ncex}
R. J. Cashmore et al., {\it Phys. Rep.} {\bf 122}, 275 (1985) ;
Y. Eisenberg et al.,
{\it Proceedings of the HERA Workshop : Physics at HERA},
W.Buchm\"{u}ller and G.Ingelman (eds), Vol 2, p.1124 (1991)
\bibitem{ccour}
T. Kon, K. Nakamura and T. Kobayashi, {\it Phys. Lett.} {\bf 233}, 461 (1989)
\bibitem{bg}
M. Drees and K. Grassie, {\it Z. Phys.} {\bf C28} (1985) 451 ;
K. Gaemers and M. Janssen, {\it Z. Phys.} {\bf C48} (1990) 491
\bibitem{susybr}
G. Altarelli et al., {\it Nucl. Phys.} {\bf B262}, 204 (1985)
\bibitem{tsutsui}
H. Tsutsui, K. Nishikawa and S. Yamada, {\it Phys. Lett.} {\bf 245}, 663 (1990)
\bibitem{elastic}
M. Drees and D. Zeppenfeld, {\it Phys. Rev.} {\bf D39}, 2536 (1989)
\bibitem{lopez}
J. L. Lopez et al., {\it Phys. Rev.} {\bf D48}, 4029 (1993)
\bibitem{lhc}
A. Bartl et al., {\it Z. Phys.} {\bf C52}, 677 (1991)
\bibitem{thomas}
T. W\"ohrmann and H. Fraas, {\it Phys. Lett.} {\bf B336}, 107 (1994) ;
{\it Phys. Rev.} {\bf D52}, 78 (1995)
\bibitem{rbel}
H. Dreiner and P. Morawitz, {\it Nucl. Phys.} {\bf B428}, 31 (1994)
\bibitem{sthera}
  T. Kon, T. Kobayashi, S. Kitamura, K. Nakamura and S. Adachi,
 {\it Z. Phys.} {\bf C61}, 239  (1994)
\bibitem{stopsg}
T. Kon, T. Kobayashi and S. Kitamura, {\it Phys. Lett.} {\bf B333}, 263 (1994)
\bibitem{stopbg}
  T. Kobayashi, T. Kon, K. Nakamura and T. Suzuki,
 {\it Mod. Phys. Lett.} {\bf A7}, 1209 (1992)
\bibitem{Schuler}
 G. A. Schuler. {\it Nucl. Phys.} {\bf  B299},  21  (1988)
\bibitem{Eichten}
E. Eichten et al., {\it Rev. Mod. Phys.} {\bf   56}, 579  (1984) ;
{\bf    58}, 1065  (1986) (Errata)
\bibitem{bases}
  S. Kawabata, {\it Comput. Phys. Commun.} {\bf 41}, 127 (1986)
\bibitem{Schrempp}
B. Schrempp, {\it Proceedings of the HERA Workshop :
Physics at HERA}, W. Buchm\"{u}ller and G. Ingelman (eds), Vol 2, p.1034 (1991)
\bibitem{Schimert}
T. Schimert, C. Burgess and X. Tata, {\it Phys. Rev.} {\bf  D32}, 707 (1985)
\bibitem{Ingelman}
 G. Ingelman, {\it Proceedings of the HERA Workshop : Physics at HERA},
 W. Buchm\"{u}ller and G. Ingelman (eds), Vol 3, p.1366 (1991)
\bibitem{IngelmanSch}
 G. Ingelman and G. A. Schuler,
 {\it Proceedings of the HERA Workshop : Physics at HERA},
 W. Buchm\"{u}ller and G. Ingelman (eds), Vol 3, p.1346 (1991)
\bibitem{Sjostrand}
T. Sj\"{o}strand, {\it Comput. Phys. Commun.} {\bf  39}, 347 (1986)
\bibitem{sgmu}
H1 Collab., DESY preprint, DESY 94-248



\end{thebibliography}
\end{document}